\documentclass[%
reprint,
superscriptaddress,
amsmath,amssymb,
aps,
longbibliography,
prb,
]{revtex4-1}

\usepackage[toc,page,header]{appendix}
\usepackage{minitoc}
\usepackage{graphicx}
\usepackage{multirow}
\usepackage{soul}


\usepackage{dcolumn}
\usepackage{bm}
\usepackage[colorlinks, linkcolor=blue, citecolor=blue, urlcolor=blue]{hyperref}

\begin{document}

\title{Quantum criticality in the nonunitary dynamics of $(2+1)$-dimensional free fermions}

\author{Qicheng Tang}
\affiliation{Zhejiang University, Hangzhou 310027, China}
\affiliation{School of Science, Westlake University, Hangzhou 310024, China}
\affiliation{Institute of Natural Sciences, Westlake Institute of Advanced Study, Hangzhou 310024, China}

\author{Xiao Chen}
\email{chenxiao.phy@gmail.com}
\affiliation{Department of Physics, Boston College, Chestnut Hill, MA 02467, USA}

\author{W. Zhu}
\email{zhuwei@westlake.edu.cn}
\affiliation{School of Science, Westlake University, Hangzhou 310024, China}
\affiliation{Institute of Natural Sciences, Westlake Institute of Advanced Study, Hangzhou 310024, China}

\date{\today}

\begin{abstract}
We explore the nonunitary dynamics of $(2+1)$-dimensional free fermions and show that the obtained steady state is critical regardless the strength of the nonunitary evolution. Numerical results indicate that the entanglement entropy has a logarithmic violation of the area-law and the mutual information between two distant regions decays as a power-law function. In particular, we provide an interpretation of these scaling behaviors in terms of a simple quasiparticle pair picture. In addition, we study the dynamics of the correlation function and demonstrate that this system has dynamical exponent $z=1$. 
We further demonstrate the dynamics of the correlation function can be well captured by a classical nonlinear master equation.
Our method opens a door to a vast number of  nonunitary  random  dynamics in free fermions and can be generalized to any dimensions.
\end{abstract}

\maketitle

\section{Introduction}
Quantum entanglement offers an information-based way to peek into quantum correlations for both ground-state static features \cite{srednicki1993entropy, HOLZHEY1994443, vidal2003entanglement, Calabrese_2004entanglement,fradkin2006entanglement, kitaev2006topological,levin2006detecting,ryu2006holographic} and nonequilibrium dynamics \cite{Calabrese_2005evolution, Calabrese_2007quantum, Kim_2013,bardarson2012unbounded,Znidaric_2008,Ho_2017,Nahum_EE_2017}.
In a closed quantum many-body system, the wave function is undergoing unitary evolution and can thermalize under its own dynamics. The steady-state reduced density matrix for a small subsystem takes a thermal form with the entanglement entropy (EE) satisfying volume-law scaling \cite{Srednicki, Deutsch, Kim_2013,Ho_2017}. This picture can be qualitatively changed if the dynamics becomes nonunitary. For instance, for
a unitary dynamics subjected to repeated local projective measurement, the thermalization process can be suppressed. As we increase the measurement rate, there can be an entanglement phase transition from a highly entangled volume-law phase to a disentangling area-law phase \cite{skinner2019measurement, li2018quantum, chan2019unitary}. This finding leads to a surge of interest in this hybrid nonunitary quantum circuit from the novel perspective of the quantum trajectory
\cite{skinner2019measurement, li2018quantum, chan2019unitary, li2019measurement, choi2019quantum, szyniszewski2019entanglement, gullans2019dynamical, tang2020measurement, bao2020theory, jian2020measurement, gullans2019scalable, zabalo2020critical, fan2020selforganized, li2020conformal, lavasani2020measurementinduced, fuji2020measurementinduced, szyniszewski2020universality, alberton2020trajectory, rossini2020measurement, lang2020entanglement, sagars2020measurementdriven, shtanko2020classical, sang2020measurement, laconis2020measurement, nahum2020measurement, turkeshi2020measurement, lavasani2020topological,li2020statistical,Cao_2019,alberton2020trajectory,lunt2020dimensional,lavasani2020topological,jian2020criticality,Nahum_EE_2017,nahum2020entanglement,turkeshi2020measurement,lunt2020dimensional,lavasani2020topological,jian2020criticality}. 
Among these developments are the quantum error correction property of the volume-law phase \cite{choi2019quantum,gullans2019dynamical,fan2020selforganized,li2020statistical}, the emergent conformal symmetry at the critical point \cite{li2020conformal}, the connection with the classical spin models \cite{nahum2020measurement, bao2020theory, jian2020measurement}, and the symmetry-protected nontrivial area-law phase \cite{lavasani2020topological,sang2020measurement}. 

A rather special situation occurs in the free-fermion system where the projective measurement-driven phase transition is absent \cite{chan2019unitary}. Different from the interacting system in which the volume-law phase is protected by the scrambling property of the unitary evolution, the EE in the free-fermion system is contributed from the nonlocal quasiparticle pairs \cite{Calabrese_2005evolution, Calabrese_2007quantum} and can be destroyed by projective measurement \cite{chan2019unitary}. Nevertheless, the absence of volume-law phase does not immediately imply the free-fermion dynamics is trivial \cite{nahum2020entanglement, jian2020criticality}. For instance, Ref.~\onlinecite{chen2020emergent} constructed a class of $(1+1)$-dimensional [$(1+1)$D] free-fermion nonunitary random dynamics and observed a stable critical phase without finely tuning the parameter. In addition, the investigation of the dynamics indicates that this model has emergent two-dimensional conformal symmetry \cite{Calabrese_2004entanglement}.

However, previous studies on the emergent quantum criticality in (non)unitary dynamics are mainly focused on the $(1+1)$D systems, less known about higher-dimensional cases.
Even for the noninteracting case, the well-established knowledge is rare.
In this paper we will explore the (non)unitary dynamics of $(2+1)$-dimensional [$(2+1)$D] free-fermion models.
The aim of this work is many-fold. First, we provide a systematical study of quantum entanglement in $(2+1)$d (non)equilibrium systems, which is complementary to the existing literature~\cite{lemonik2016entanglement, cotler2016entanglement, zhao2019logarithmic}.
Second, we investigate the nonunitary random dynamics of $(2+1)$D systems, and explore the emergent quantum criticality and its universal entanglement features.

By performing large-scale numerical simulation, we find that for most of the cases, the EE has a logarithmic violation of the area law, the same as the system with finite Fermi surface. 
We further study the mutual information (MI) and the squared correlation function of the steady state and we find that they all have the power-law scaling form, another signature of the criticality of the steady-state wave function.
Notably, our generating steady state is distinguished from the ground state of the quantum metal due to the absence of the Fermi surface.
Understanding the emergent critical entanglement scaling in our model requires knowledge beyond the Widom conjecture~\cite{Widom1982, wolf2006violation, gioev2006entanglement}.

In this work, two analytical approaches are provided to understand the critical behavior observed in our systems. In the first approach, we assume that the EE in this free-fermion dynamics comes from the quasiparticle pairs \cite{nahum2020measurement}. By defining the  quasiparticle pair probability distribution, we successfully explain both the scaling behaviors of the EE and MI. In the second approach, we introduce a Brownian nonunitary free-fermion model \cite{chen2020emergent} and derive a master equation for the squared correlation which gives an intuitive picture for the spreading of the correlation function and reproduces all the interesting results found in the numerics of the discrete model. Most importantly, this method implies that the dynamical exponent $z=1$ which is also confirmed in the discrete dynamics. 

This paper is organized as follows. In Sec.~\ref{sec:model}, we introduce the model of nonunitary dynamics, and the method used for computation. 
In Sec.~\ref{sec:unitary}, we study the unitary dynamics of $(2+1)$D free fermion, which provides a generic picture about volume-law behavior of unitary dynamics. 
In Sec.~\ref{sec:non_unitary}, we investigate the nonunitary dynamics of $(2+1)$D free fermion. We present numerical results and provide a detailed analysis based on two theoretical methods. At last, we give a summary in Sec.~\ref{sec:summary}.

\begin{figure}\centering
	\includegraphics[width=0.85\columnwidth]{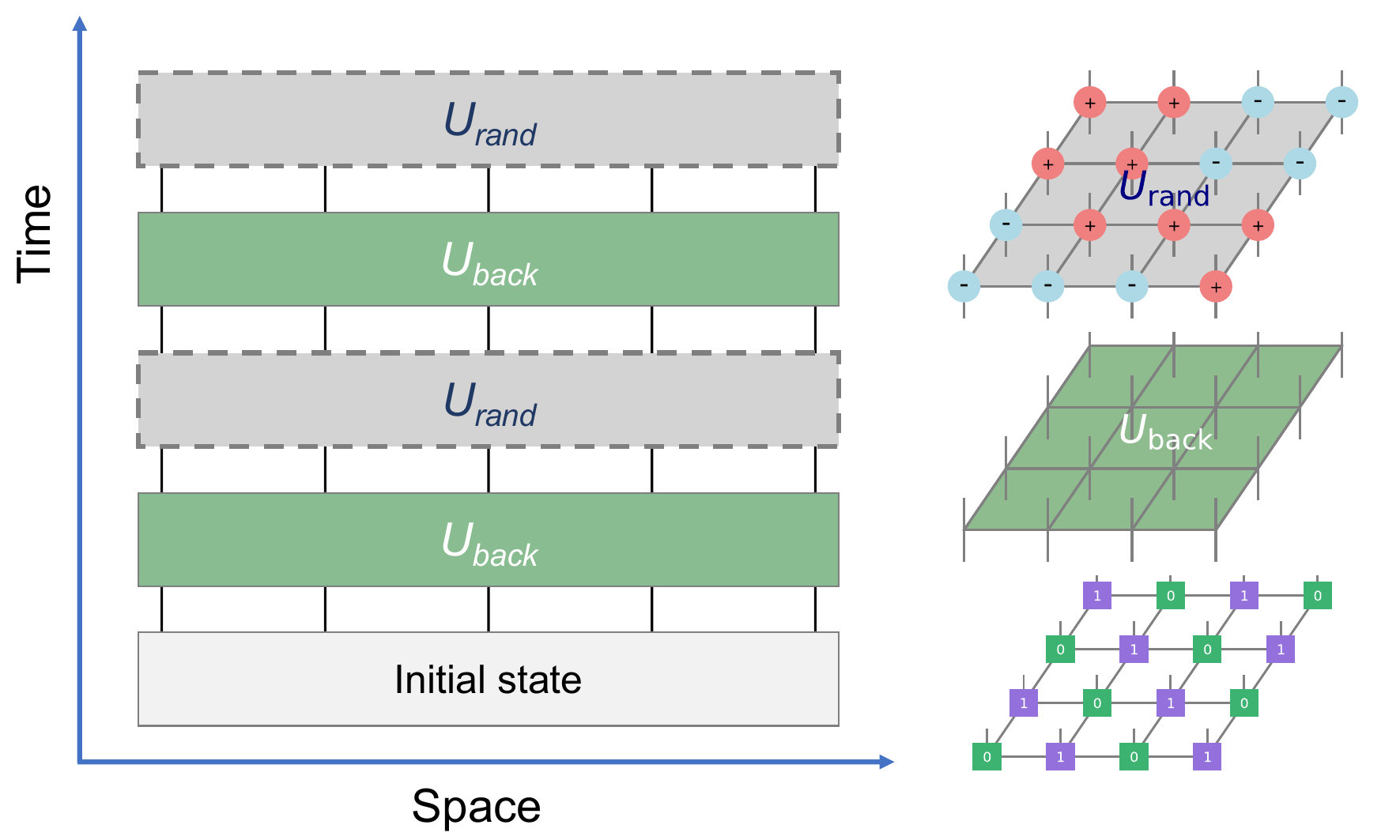}
	\caption{
		\label{fig:schematics}
		(Left panel) Schematic of mixed nonunitary random dynamics of $(2+1)$D free fermion model,
		under the unitary background time evolution represented by $U_{\rm{back}}$
		and imaginary random time evolution represented by $U_{\rm{rand}}$.
		(Right panel) The initial state prepared to be a half-filling product state (bottom),
		one period of the unitary background time evolution driven by a (homogeneous) tight-binding model (middle),
		and the imaginary time evolution driven by random onsite potentials (top).
	}
\end{figure}

\section{Model and method}\label{sec:model}
We consider the discrete nonunitary random dynamics described by $U(T)=\prod_{t=1}^T U(t)$ (see Fig.~\ref{fig:schematics}). At time step $t$, $U(t)$ is composed of both unitary and imaginary time evolution, i.e., 
\begin{align}\label{eq:discrete}
U(t)= U_{\rm{rand}}(\beta) U_{\rm{back}}(\tau) ,
\end{align}
where the unitary background dynamics $U_{\rm{back}}(\tau) = \exp[{-i\tau  H_{\rm{TB}}}]$ is governed by a  $(2+1)$D tight-binding lattice model with 
$H_{\rm{TB}} = - \sum_{\left\langle i, j \right\rangle} c^{\dagger}_i c_j$
($\left\langle i, j \right\rangle$ represents the neighboring pairs on the lattice and strength of the hopping is set to be unit). The random imaginary dynamics is $U_{\rm{rand}}(\beta)=\exp[{- \beta H_{\rm{rand}} }]$ and $H_{\rm{rand}}$ denotes a simple random onsite potential 
$H_{\rm{rand}} = \sum_i \mu_i c^{\dagger}_i c_i$.
$\mu_i$ is a random number in both spatial and temporal directions, and it takes a simple two-component distribution and has half-probability to be $+1$ and half-probability to be $-1$. In $U(t)$, $\beta$ and $\tau$ represent the strength of the imaginary random potential and the unitary  dynamics, respectively.
For simplicity, we fix $\tau=1$ and only leave $\beta$ as a tuning parameter.

We are interested in the wave-function dynamics,
\begin{align}
|\psi(T)\rangle=\frac{1}{\sqrt{Z(T)}}U(T)|\psi_0\rangle ,
\end{align}
where $Z(T)=\langle \psi_0|U^\dag(T)U(T)|\psi_0\rangle$. In this paper, we are mainly focused on the initial state $|\psi_0\rangle$ as a product state. Under the nonunitary dynamics, $|\psi(T)\rangle$ remains a fermionic Gaussian state and therefore the information of the wave function is fully encoded in the two-point correlation matrix $C(T)$ \cite{bravyi2004lagrangian,chen2020emergent} with
$C_{ij}(T)=\langle \psi(T)|c_i^{\dag}c_j|\psi(T)\rangle$.
From the $C(T)$ matrix, we can further compute the von Neumann EE \cite{chung2001density, Peschel_2003calculation, Peschel_2009reduced}
\begin{align}
S_{\rm{vN}}=-\mbox{Tr}[C_A\ln C_A+(1-C_A)\ln (1-C_A)],
\end{align}
where $C_A$ is the correlation matrix for subsystem $A$. 

To accelerate the computations and reach larger system sizes,
the calculation of matrix-matrix multiplication and eigenvalue decomposition is performed in parallel by GPU on Nvidia V100 systems.  
This allows us to reach the largest system size up to $L\times L = 176 \times 176$ on the square lattice. 
We also carefully confirm the choice of simulation parameters, such as $\beta$ and the filling factor $\nu$, do not influence the qualitative picture shown below (see details in the Appendices). Since we study the random dynamics, all the EE and correlation function results are obtained through ensemble averaging over different circuit realizations.

\section{Unitary dynamics}\label{sec:unitary}
\begin{figure}\centering
	\includegraphics[width=\columnwidth]{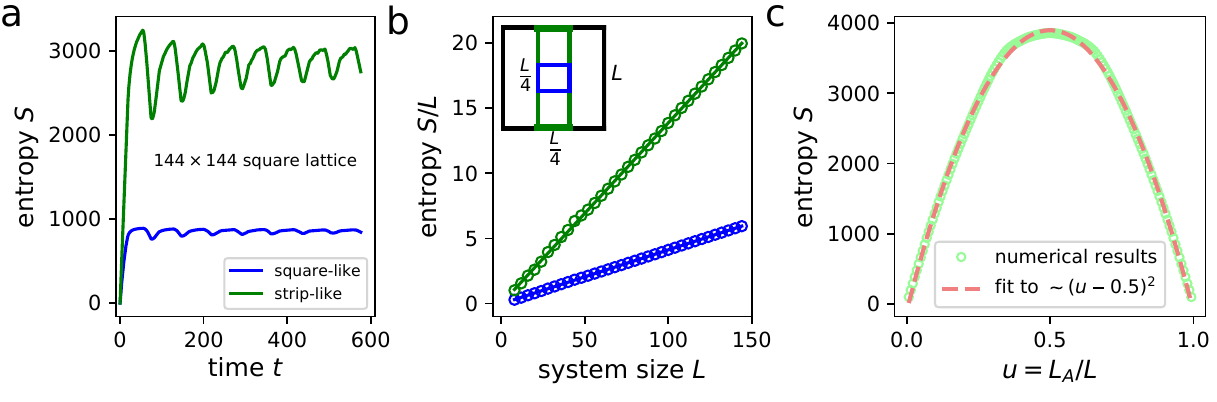}
	\caption{
		\label{fig:Scaling_RealEvoEE_square}
		\textbf{Unitary dynamics of EE.}
		(a) The time evolution of EE $S(t)$ with (sub)system configurations shown in the insert of (b).
		(b) The late time EE per length $S(t\to \infty)/L$ with (sub)system configurations shown in the insert. 
		Here the lines show linear fitting in form $S/L = aL + b$ for large $L$, which indicates to a volume law $S \sim L^2$.
		(c) The late time EE $S(t\to \infty)$ as a function of $u=L_A/L$ for $L_A \times L$ strip-like subsystem (see the green strip in the insert of (b)) with fixed $L=144$. 
		Numerically we find the approximate scaling form of $S \sim (u-0.5)^2$.
		For simplicity the initial state is chosen to have N\`eel-type order.
		Here we present results for the square lattice; see other lattices in Appendix~\ref{app:unitary}).
	}
\end{figure}

We start by discussing the unitary time evolution of $(2+1)$D free-fermion lattice model (without nonunitary terms $U_{\rm rand}$ in Fig.~\ref{fig:schematics}).
In $(1+1)$D, it is generally expected that quenching a short-correlated state governed by a free-fermion Hamiltonian produces a volume-law EE at large time due to the quasiparticle picture ~\cite{Calabrese_2005evolution, Calabrese_2007quantum}.
This conclusion is expected to hold also in higher dimensions.
In Fig.~\ref{fig:Scaling_RealEvoEE_square}, we present numerical results of EE during unitary dynamics of $(2+1)$D free fermions.
We consider EE for subsystem being $L_A \times L_A$ square or $L_A \times L$ cylinder defined on the $L \times L$ square lattice with periodic boundary condition (torus geometry).
As shown in Fig.~\ref{fig:Scaling_RealEvoEE_square} (a), during unitary dynamics, the EE grows rapidly and saturates to a volume-law scaling with some oscillations, a strong evidence that the entanglement is caused by the propagating quasiparticle pairs.

We examine the volume-law entanglement scaling in different geometries. 
The total system is a square torus and the subsystem is either a cylinder or a square (see Fig.~\ref{fig:Scaling_RealEvoEE_square} (b)). We fix the ratio $u = L_A/L = 1/4$. By changing the total system length $L$, we observe that in both cases, EE scales as $L^2$. 
These scaling behaviors indicate a volume-law EE as expected.
Moreover, we have also considered honeycomb and Lieb lattices (see details in Appendix~\ref{app:unitary}). Both of them display very similar scaling, showing that the produced volume law is universal for different lattices.

In integrable models, the entanglement dynamics can be understood by the \emph{quasiparticle} picture~\cite{Calabrese_2005evolution, Calabrese_2007quantum}.
The basic idea is to consider the initial state as the source of quasiparticles (entangled pairs), and the entanglement growth as the consequence of their propagation that is driven by the quenching Hamiltonian.
The entangled pairs move with certain group velocity (the explicit form relies on the lattice details) into distant sub-regions (say) $A$ and $B$, so that the entanglement region is growing during the dynamics.
We numerically confirmed this picture by measuring the MI (defined as $I_{A, B} = S_A + S_B - S_{A \cup B}$), which exhibits a clear wave-front in space-time as a signature of quasiparticle propagation (see details in Appendix~\ref{app:unitary})), akin to the $(1+1)$D systems.

\section{Nonunitary dynamics}\label{sec:non_unitary}
In this section, we turn to explore the nonunitary quantum dynamics of $(2+1)$D free fermions.
It is found that the previously presented results in Sec.~\ref{sec:unitary} are dramatically changed by introducing additional imaginary random potentials.
We first present numerical results and then discuss the underlying physical picture via two theoretical methods.

\subsection{Numerical results}\label{sec:non_unitary_EE}
\begin{figure}\centering
	\includegraphics[width=\columnwidth]{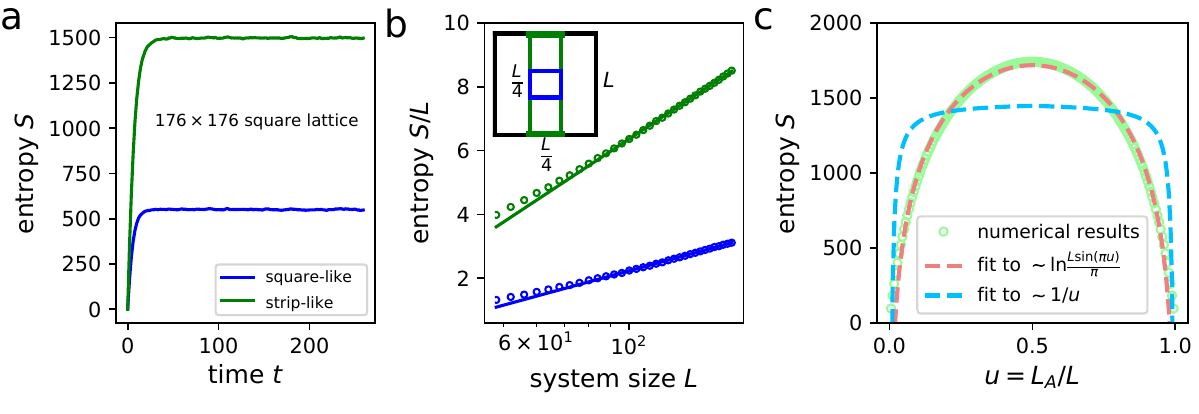}
	\caption{
		\label{fig:Scaling_NonUnitaryEE_square}
		\textbf{Mixed nonunitary dynamics of EE.}
		(a) Time evolution of EE $S(t)$ with (sub)system configurations shown in the insert of (b).
		(b) The late time EE per length $S(t\to \infty)/L$ with (sub)system configurations shown in the insert.
		The $x$ axis is set to be in log-scale to perform a linear fitting in form $S/L = a\log L + b$ for large $L$, which indicates to a logarithmic violation of area-law $S \sim L\log L$.
		(c) The late time EE $S(t\to \infty)$ as a function of $u = L_A/L$ for $L_A \times L$ strip-like subsystem (see the green strip in the insert of (b)) with fixed $L=176$.
		Here we present results for square lattice with $\beta=1$ and $\nu=0.5$; see other lattices(settings) in Appendix~\ref{app:mixedEE}.
	}
\end{figure}

In Fig.~\ref{fig:Scaling_NonUnitaryEE_square}, we show the numerical results of the EE in our model of the mixed nonunitary dynamics on square lattice, and the (sub)system configurations are considered to be the same as in Fig.~\ref{fig:Scaling_RealEvoEE_square}.
The main feature is that the dynamics leads to a steady state with nontrivial entanglement structure.
We find this steady state is quite stable, with very small fluctuations of the EE as a function of time.
In addition, the very similar behaviors are observed on different lattice geometries or electron filling numbers (see details in Appendix~\ref{app:mixedEE})), which reflects the observed behavior is quite robust, intrinsic to the nonunitary random process.

We expect the steady state generated by the mixed nonunitary dynamics to be critical, as the previous observation in the $(1+1)$D case~\cite{chen2020emergent}.
However, for $(2+1)$D critical systems, the possible entanglement scaling forms are not constrained to a simple form as in $(1+1)$D.
On the one hand, for a large class of the critical systems, e.g., the gapless Dirac cone which belongs to the $(2+1)$D conformal field theory (CFT), 
the EE satisfies area-law scaling with a universal subleading term depending on the geometry of the total system and the subsystem~\cite{fradkin2006entanglement, Casini_2007, Casini_2010, Chen_2015}.
For torus geometry it has been found that this subleading term scales as $\propto 1/u$ for $u \to 0$ and $\propto 1/(1-u)$ for $u \to 1$, which is symmetric around $u = 0.5$.~\cite{Chen_2015}
On the other hand, it is also known that in the fermionic system with a finite Fermi surface, the EE has a logarithmic violation of the area law~\cite{wolf2006violation, gioev2006entanglement}.
This can be understood by considering the EE of the $(2+1)$D system as summation of the EE for each $(1+1)$D gapless modes with $S_{\rm 1D} \propto \log L$.
It is natural to apply this idea to the torus geometry and the EE for a cylinder has $S \propto L\log[L\sin(\pi u)]$.
These scaling forms are typical for $(2+1)$D critical systems, and we will consider them as the guideline of the present investigation.

As shown in Fig.~\ref{fig:Scaling_NonUnitaryEE_square} (b), we present the results for both square- and strip-like subsystems with fixed ratio $u$, the leading term in the EE scales as 
\begin{align}
	S_{\rm{vN}} \sim a L\log L .
\end{align}
Additionally, in Fig.~\ref{fig:Scaling_NonUnitaryEE_square} (c), we further study the scaling behavior of the two-cylinder EE by varying the cylinder length $L_A$.
The result indicates that the two-cylinder EE takes the form 
\footnote{Numerically, we find that the R\'enyi EE takes the same form with the coefficient $a\sim (1+1/n)$.}
\begin{align}
	S_{\rm{vN}}=a L_y\log \frac{L_x\sin(\pi u)}{\pi}+b .
\end{align}
The same scaling behavior can also be obtained in $L_y$ copies of the $(1+1)$D critical systems, as discussed above.
In a word, for all of the (sub)system configurations, we find evidence that the steady-state EE has a logarithmic violation of area law, akin to the ground state of a Fermi liquid.

\begin{figure}\centering
	\includegraphics[width=\columnwidth]{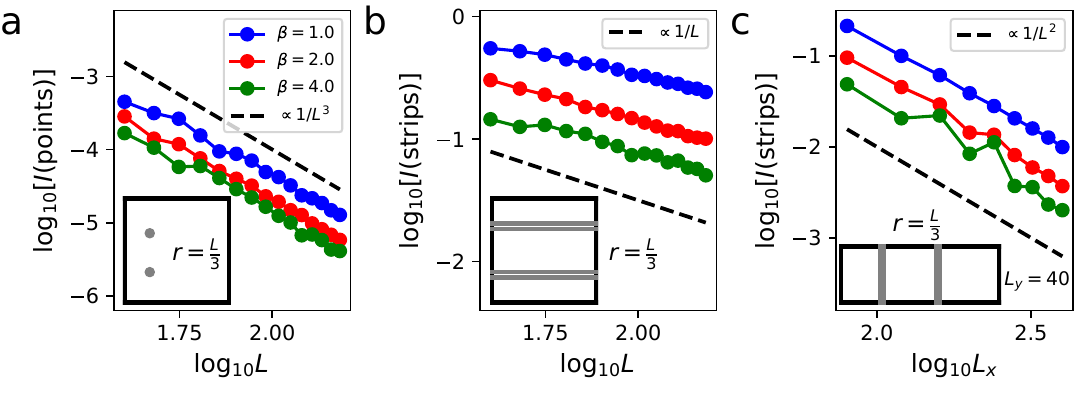}
	\caption{
		\label{fig:Scaling_NonUnitaryMutinf_square}
		\textbf{Scaling behavior of MI for mixed nonunitary steady state.}
		(a) MI between two points $I({\rm{points}})$ on $L \times L$ torus, simulated with total system size up to $L = 176$.
		(b) MI between two strips $I({\rm{strip}})$ on $L \times L$ torus, simulated with total system size up to $L = 176$.
		(c) MI between two strips $I({\rm{strip}})$ of steady state on $L_x \times L_y$ thin torus (with fixed $L_y=40$), simulated with total system size up to $L_x = 400$.
		The distance between two subsystems is set to be $r = \frac{L}{3}$, and the geometry of total system and subsystem is plotted in the insert.
		Here we present results for square lattice; see other lattices in Appendix~\ref{app:mixedMI}.
	}
\end{figure}

The above results suggest that the steady state is critical. We further study the MI to better understand the entanglement structure of this steady state.
When $A$ and $B$ are two distant small subsystems, in the critical wave function, we expect that $I(A,B)\sim 1/r^\alpha$, where $r$ is the distance between $A$ and $B$. To extract the critical exponent $\alpha$ numerically in a finite system, we take $A$ and $B$ as single points and fix the ratio $r/L$. As shown in Fig.~\ref{fig:Scaling_NonUnitaryMutinf_square} (a), we vary $L$ and find that 
\begin{align}
	I(A,B)\sim 1/L^{3},
\end{align} 
indicating that $\alpha=3$. This is because in the finite system we expect that $I\sim  f(r/L)/L^{\alpha}$, in which the scaling function $f(r/L)\sim (L/r)^{\alpha}$ in the limit $r\ll L$. Moreover, we compute the MI between two strips (Fig.~\ref{fig:Scaling_NonUnitaryMutinf_square}(b) and \ref{fig:Scaling_NonUnitaryMutinf_square}(c)) and we find MI exhibits power-law decay behavior with  $I(A,B)\sim L_y/ r^2$.  
(The reason will be clarified later in Sec. IV B.)

We further explore the scaling of the correlation functions. Since this is a random system, the averaged correlation function $\overline{C_{i,i+r}}=0$ for $r>0$. However, when $r\gg 1$, the averaged squared correlation function is nonzero and has $|C(r)|^2\equiv \overline{|C_{i,i+r}|^2}\sim 1/r^3$, the same as the MI between two distant points. 

Additionally, by analyzing the time dynamics of $|C(r,T)|^2$, we find that it satisfies the form
\begin{align}\label{eq:corr}
	|C(r,T)|^2\sim \frac{F(r/T)}{T^3}.
\end{align}
This result is confirmed by the data collapse in Fig.~\ref{fig:Scaling_NonUnitaryCorrelation_square}. In particular, when $r\gg T$, the scaling function $F(r/T)\sim \exp(-c r/T)$  and when $r\ll T$, $F(r/T)\sim T^3/r^3$. Therefore at early time, $|C(r,T)|^2$ decays exponentially in $r$ with a correlation length proportional to $T$. At late time, $|C(r,T)|^2$ reproduces the $1/r^3$ power-law decay. The existence of the scaling function $F(r/T)$ indicates that the dynamical exponent $z=1$, again the same as the $(1+1)$D nonunitary random dynamics.

\subsection{Nonlinear master equation}

\begin{figure}\centering
	\includegraphics[width=0.95\columnwidth]{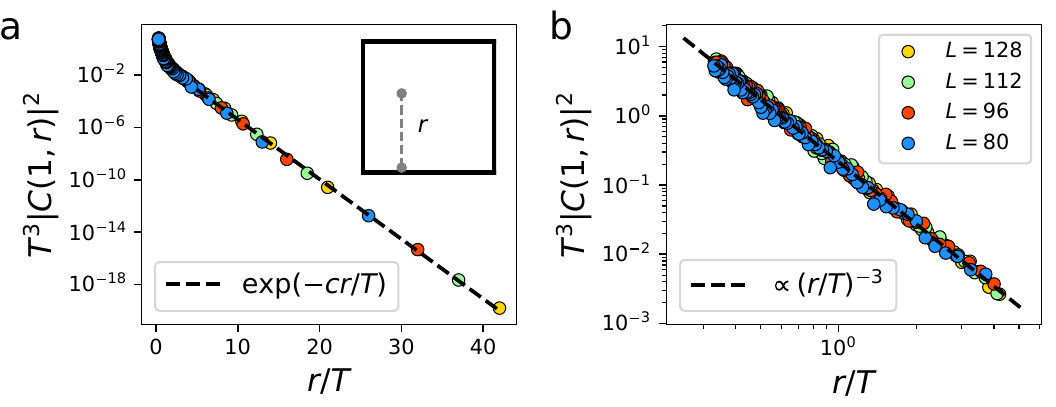}
	\caption{
		\label{fig:Scaling_NonUnitaryCorrelation_square}
		\textbf{Mixed nonunitary dynamics of correlation function $\bm{|C(1,r)|^2}$}.
		(a) Semi-log plot of $T^2|C(1,r)|^2$ at the limit $T \ll r$. Insert shows the considered two points with distance $r$. 
		The dash line represents a linear fitting to the form of $\log T^3 |C(1,r)|^2 = -c r/T + b$, which indicates an exponential decay of $|C(1,r)|^2 \sim T^{-3}\exp(-cr/T)$.
		(b) Log-log plot of later dynamics (before reaching steadiness) of $T^2|C(1,r)|^2$.
		The dashed line represents a linear fitting to the form of $\log T^3 |C(1,r)|^2 = -a \log(r/T) + b$ with $a=3$, which indicates a power-law decay of $|C(1,r)|^2 \sim r^3$.
		Here we present results for square lattice with $\beta=1$ and $\nu=0.5$, see other settings in Appendix~\ref{app:mixedCorr}.
	}
\end{figure}

To understand the  correlation function dynamics in Eq.~\eqref{eq:corr}, we follow the method introduced in Ref.~\onlinecite{chen2020emergent} and derive a master equation to describe the evolution of the squared correlation function.
We consider a Brownian nonunitary dynamics, and 
derive a master equation for averaged distribution function $f_{w,z}=\sum_{x,y}|C_{(x,y),(x+w,y+z)}|^2/L^2$ (see details of this derivation in Appendix~\ref{app:master}).
In the spatially continuous limit, we have
\begin{align}
	\label{eq:master}
	\partial_t f_{w,z} =\ & \theta \nabla^2 f_{w,z} +\mu \delta_{(w,z),(0,0)} \nonumber\\
	&-\kappa f_{w,z} + \int_{-\infty}^\infty  f_{x, y} f_{w-x, z-y} dx dy ,
\end{align}
where $\mu$, $\kappa$ are both positive $\mathcal{O}(1)$ constants and $\theta \sim {\tau}/{\beta}$ is the ratio between the strength of unitary and imaginary time evolution.
This equation is not exact, but does take the scaling form described in Eq.~\eqref{eq:corr}. 
The diffusion term comes from the unitary dynamics, the source term is due to the diagonal element in the correlation matrix, and the nonlinear convolution term is caused by the nonunitary evolution. 
The latter is important and is responsible for the interesting scaling behavior found in Fig.~\ref{fig:Scaling_NonUnitaryCorrelation_square}.  
We numerically solve the discrete version of Eq.~\eqref{eq:master} and we find that
\begin{align}
	f_{w,z} \sim
	\begin{cases}
		t^{-3} \exp(-cr/t), & t \ll r \\
		1 / r^3, & t \gg r
	\end{cases}
\end{align}
with Euclidean distance $r = \sqrt{w^2+z^2}$.
Both early and late time results are consistent with the numerics, thus we conclude that the nonlinear master equation captures the scaling behavior of evolved correlations during mixed nonunitary dynamics. 
Moreover, numerically we find that the nonlinear master equation with different values of $\theta$ leads to universal scaling behavior of $f_{w,z}$, consistent with the argument made above.
This is because the diffusive dynamics is much slower than the aggregation dynamics caused by the convolution term.
The insensitivity of $\theta$ implies that a unitary chaotic background dynamics is not necessary for accessing the nontrivial entanglement structure observed in our designed model. This important fact leads to the further exploration on other types of nonunitary dynamics, where the very similar behavior is observed (see Sec.~\ref{sec:PureImag}).
Additionally, if we ignore the diffusive term from unitary dynamics by setting $\theta = 0$ and assume steadiness of $f_{w,z}$ ($\partial_t f_{w,z}=0$), the ansatz for solving the equation can be obtained analytically as $f(r) \sim 1/r^3$, which is consistent with the numerical solution (see details in Appendix~\ref{app:master}).

\subsection{Quasiparticle picture} 
In  the steady state, if we assume that the entanglement is mainly contributed by the quasiparticle pairs, we can compute the distribution of the quasiparticle pairs and estimate the EE. 
The MI between two subsystems $A$ and $B$ can be written as
\begin{align}\label{eq:quasi-particle}
	I_{A,B} \sim \int_A dV_A \int_B dV_B P(r_{A,B}) .
\end{align}
Since the MI between two points scale as $1/r^3$, we expect the probability $P(r)$ of a quasiparticle pair separated by distance $r$ scales as $1/r^3$. 
If $A$ is a disc/strip and $B$ is the complement of the system,  we analytically obtain the EE $S_A \sim L\ln L$   (see derivation in Appendix~\ref{app:quasi}), which is consistent with our numerics of the mixed nonunitary dynamics.
Moreover, MI between two strips takes the scaling form $I({\rm{strips}}) \propto  L_y/r^2$, also consistent with our numerical simulation.

Our quasiparticle picture is inspired by Ref.~\onlinecite{nahum2020entanglement}, which considered a measurement-only free-fermion dynamics. The steady state in their model is composed of quasiparticle pairs and has the same entanglement and MI scaling. Two quasiparticle pairs can combine into new quasiparticle pairs due to the measurement. Intuitively, this aggregation process is analogous to the convolution term in our master equation for the Brownian circuit. It would be interesting to better understand the connection between these two models in the future. 

Notice that although the steady state in the nonunitary random dynamics and the ground state of the quantum metal with a finite Fermi surface take the same entanglement scaling, these two states are quite different and can be distinguished by the correlation function/MI. For the latter,  the correlation function in real space is the Fourier transform of occupation number $\langle n_k\rangle$ in the momentum space and depends on the shape of the Fermi surface \cite{swingle2012renyi}. 
However, in the nonunitary dynamics, due to the stochastic randomness, the quantum coherence is lost and the Fermi surface is absent (see Appendix~\ref{app:absence}). Presumably the correlation is  contributed by the quasiparticle pairs and the dynamics can be described by a classical master equation.

\subsection{Random imaginary evolution}\label{sec:PureImag}

\begin{figure}\centering
	\includegraphics[width=0.9\columnwidth]{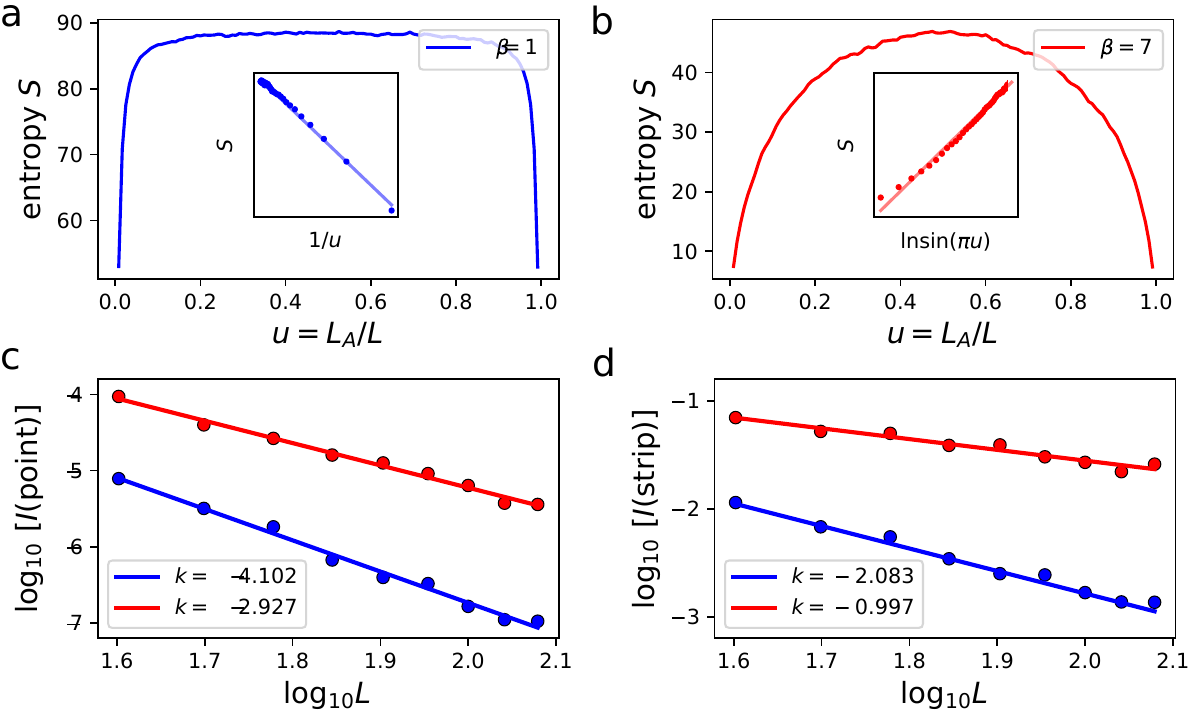}
	\caption{
		\label{fig:Scaling_PureImag_RandOnsite}
		\textbf{Entanglement in pure imaginary random dynamics.}
		(a),(b) The late time EE $S(t\to \infty)$ as a function of $u=L_A/L$ for $[0, L_A] \times [0,L]$ strip-like cut-subsystems, calculated with (blue) $\beta = 1$ and (red) $\beta = 7$. (Insets) Scaling of $S(t\to \infty)$ with appreciate functional form.
		MI (c) between two points $I({\rm{points}})$, and (d) between two strips $I({\rm{strip}})$.
		The lines in (c) and (d) are linear fittings to the form of $\log I = k\log L + b $, which indicates a power-law decay as $I \propto L^k$.
	}
\end{figure}

We further investigate the case of discrete random imaginary dynamics as an extension of our reported mixed nonunitary dynamics.
We consider the discrete time evolution that is driven by $U(T) = \prod_{t=1}^{T} U(t)$ with
\begin{align}
	U(t) = \exp[-H_{\rm TB}] \exp[-\beta H_{\rm rand}] ,
\end{align}
where  $H_{\rm{TB}}$ and $H_{\rm{rand}}$ are defined as before.
When $\beta=0$, the steady state is the ground state of $H_{\rm{TB}}$. 
We consider $H_{\rm{TB}}$ on the honeycomb lattice, where the ground state is a Dirac semi-metal with the EE satisfying the area-law scaling \cite{Casini_2007,Casini_2010}. 
As we have discussed in Sec.~\ref{sec:non_unitary_EE}, for the two-strip EE defined on the torus, there is a universal subleading term which scales as $1/u=L/L_A$ when $u\ll 1$. 
When $\beta>0$, we observe that the steady state of the random imaginary dynamics remains critical. In particular, we observe two distinct EE scaling behaviors as we vary $\beta$. When $\beta$ is small, the steady state satisfies the area-law scaling with a subleading $1/u$ correction term. The MI $I({\rm{points}})$ (also the squared correlation function) $\sim 1/r^4$ (See Fig.~\ref{fig:Scaling_PureImag_RandOnsite}). These results are the same as the $\beta=0$ limit, indicating that the Dirac cone in $(2+1)$D is stable under weak disorder perturbation \cite{Harris_1974}. When disorder is strong, the steady state logarithmically violates area-law  scaling with $I({\rm{points}}) \sim 1/r^3$, akin to the mixed nonunitary random dynamics (see Fig.~\ref{fig:Scaling_PureImag_RandOnsite}). More detailed numerical simulation indicates that the transition occurs at $\beta_c \simeq 5$. We also consider $H_{\rm{TB}}$ on the square lattice which has a finite Fermi surface. In this case, when $\beta$ is finite, we observe that the steady state has $\sim 1/r^3$ quantum correlations and the EE logarithmically violates area-law scaling.

\section{Conclusion and discussion}\label{sec:summary}
We have presented a thorough investigation of nonunitary random free-fermion dynamics in $(2+1)$D.
A protocol, based on mixed nonunitary dynamics, is designed to generate the critical steady states, evidenced by a typical entanglement entropy scaling of $S \sim L\log L$ and mutual information scaling of $I({\rm{points}}) \sim 1/r^3$.
These scaling behaviors are found to be universal on different lattice geometries, and robust by varying parameters.
To uncover the nature of criticality and the entanglement structure of the steady state, we exploit a combination of nonlinear master equation and a physical picture based on quasiparticle pairs dynamics. 

We further investigate the stability of Dirac semi-metal and metal with finite Fermi surface subjected to imaginary random perturbation. We find that when the randomness is strong enough, they all flow to the same critical states found in the mixed nonunitary dynamics. It would be interesting to introduce interaction in the random free-fermion dynamics and examine the stability of the critical phase. We leave this for the future study.

Remarkably, our results can be easily generalized into higher spatial dimensions (see Appendix~\ref{app:higher}). A simple scaling analysis of the nonlinear master equation with a convolution-like term leads to $f(r) \sim 1/r^{d+1}$. 
Taking this value into the integral of Eq.~\eqref{eq:quasi-particle}, it gives rise to $S \sim L^{d-1}\ln L$ for any spatial dimensions.

\begin{acknowledgments}
We thank Adam Nahum and Chao-Ming Jian for fruitful discussion.
Q.T. and W.Z. are supported by the foundation of Westlake University, by the Key R$\&$D Program of Zhejiang Province, China (Grant No. 2021C01002).  X.C. thanks Matthew
Fisher, Yaodong Li, and Andrew Lucas for collaborations on related projects. 
\end{acknowledgments}



\appendix
\renewcommand\thefigure{S\arabic{figure}}
\setcounter{figure}{0} 

\renewcommand\thetable{S\arabic{table}}

\begin{widetext}
	\vspace{4mm}
	\begin{center}
		\textbf{Appendices for: ``Quantum criticality in the nonunitary dynamics of $(2+1)$D free fermions''}
	\end{center}
	\vspace{1mm}
	
In the Appendices we present details related to entanglement dynamics of $(2+1)$D free fermion, and more supporting data.
First of all, we show the entanglement scaling behavior of ground state and steady state of unitary dynamics on different lattice geometries. Although these results are conventional and part of them are well known, here we present a systematical summary that is convenient for searching of them.
Second, we test the reported mixed nonunitary random dynamics in the main text on different lattice geometries and with various settings of simulating parameters.	The emergent criticality is found to be robust and the scaling behavior of quantum entanglement still holds.
Third, we propose a possible finite-size scaling form for mutual information that is universal for different lattice geometries. Although it cannot be derived analytically, we find that the data collapse of the numerical data supports the conjectured form.
Fourth, we investigate the occupation number in momentum space $n(k)$ as an indicator of the absence of a Fermi surface. This result provides strong evidence that the formation of the logarithmic violation of area-law entanglement has different origin with Fermi liquids.
At last, we present the details of the analytical approach towards the emergent criticality, including the nonlinear master equation and the quasiparticle picture.

\vspace{1mm}

\end{widetext}


\begin{table*}
	\caption{\label{tab:table1}A brief summary of the entanglement scaling forms for ground state and steady state at half-filling during (non)unitary dynamics on different lattices. Here we use $u={L_A}/{L}$ to represent the ratio of the subsystem with length $L_A$ cutting from a $L\times L$ lattice.}
	\begin{ruledtabular}
		\begin{tabular}{ccccc}
			\\ [-2 ex]
			&Square&Honeycomb&Lieb&\\ [-2 ex]
			\\ \hline
			\\[-2.5 ex]
			Ground State ($L_A\times L_A$ sub-square, fixed $u$, changing $L$)&$S \sim L \log L$&$S \sim L$&$S \sim L^2$ \\
			Ground State ($L_A\times L$ sub-strip, fixed $u$, changing $L$)&$S \sim L \log L$&$S \sim L$&$S \sim L^2$ \\
			Ground State ($L_A\times L$ sub-strip, changing $u$, fixed $L$)&$S\sim \log \sin(\pi u)$&$S\sim 1/u$&$S \sim u^2$ \\
			Unitary Dynamics&$S \sim L^2$&$S \sim L^2$&$S \sim L^2$ \\
			Mixed Nonunitary Dynamics&$S \sim L \log L$&$S \sim L \log L$&$S \sim L \log L$ \\
			Random Imaginary Dynamics (weak randomness)&$S\sim L\log L$&$S\sim L$& -- \\
			Random Imaginary Dynamics (strong randomness)&$S\sim L\log L$&$S\sim L\log L$& -- \\
			\\[-2.5 ex]
		\end{tabular}
	\end{ruledtabular}
\end{table*}

\section{Summary of $(2+1)$D scaling behaviors}
In this appendix, we first show the entanglement scaling behavior of the ground state, and then move to the steady state of unitary dynamics.
Various lattice geometries, including square, honeycomb, and Lieb lattices, are considered.
Although these results are conventional and part of them are well known, here we present a systematical summary that is convenient for searching of them.
Please note that, for unitary dynamics and mixed nonunitary dynamics, the scaling form is universal.

\subsection{Scaling of entanglement entropy of the ground state on different lattices}

The ground-state properties are not only important for studying static behaviors, but also are illuminating for investigation on out-of-equilibrium processes.
In this section, we present the ground-state entanglement scaling of the tight-binding model on square, honeycomb, and Lieb lattices.
The three considered lattices have very different energy dispersion and so that their ground states exhibit various entanglement scaling behaviors.
We investigate the scaling behavior under different (sub)system-cut configurations, including changing and fixing the ratio of subsystem.

\begin{figure}\centering
	\includegraphics[width=\columnwidth]{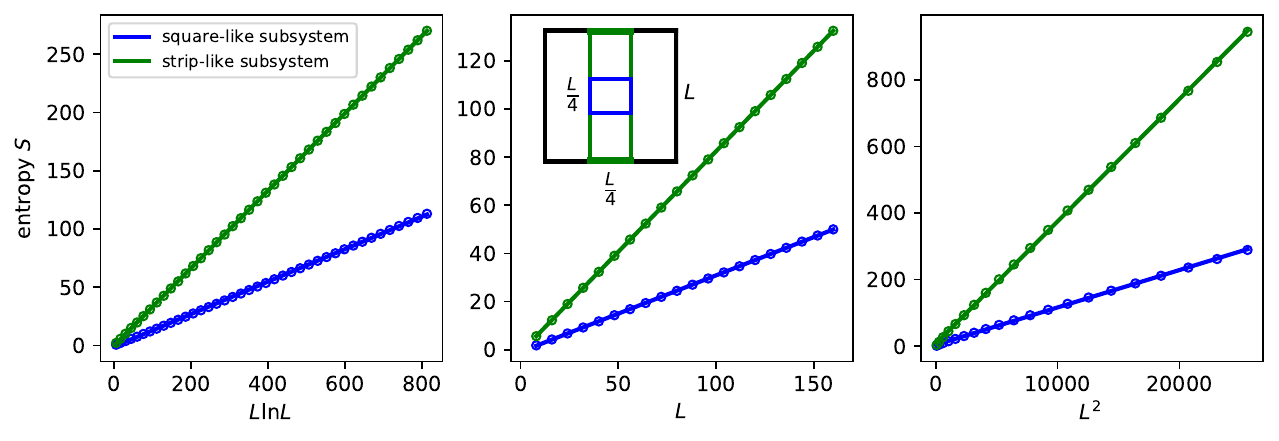}
	\caption{
		\label{fig:GS_EE_scaling}
		The ground-state entanglement entropy of the homogeneous tight-binding model on (left) square, (middle) honeycomb, and (right) Lieb lattices at half-filling. 
		As shown in the inset, we consider two kinds of subsystem configurations: $\frac{L}{4} \times \frac{L}{4}$ square-like subsystem (blue dots) and $\frac{L}{4} \times L$ strip-like subsystem (green dots). 
		Here the total system size is simulated up to $160 \times 160$.
	}
\end{figure}

\begin{figure}\centering
	\includegraphics[width=\columnwidth]{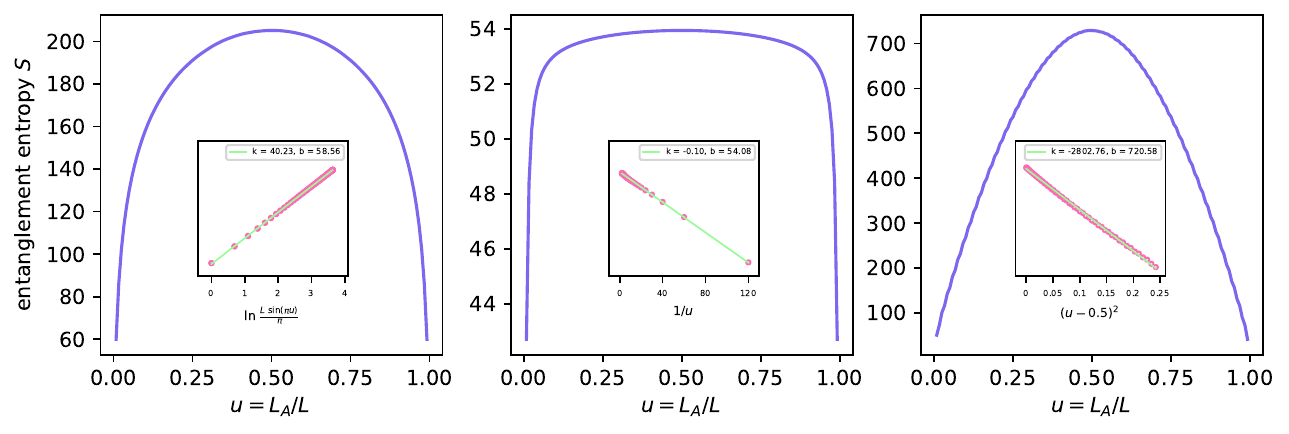}
	\caption{
		\label{fig:GS_EE_fix}
		The ground-state entanglement entropy of the homogeneous tight-binding model on (left) square, (middle) honeycomb, and (right) Lieb lattices at half-filling. The cut-subsystem configuration is set to $[0, L_A] \times [0, L]$ on a $L\times L$ lattice. Here the total system size is fixed to be $L=120$.
	}
\end{figure}

In Fig.~\ref{fig:GS_EE_scaling}, we plot the ground-state entanglement entropy on $L \times L$ lattices, with $\frac{L}{4} \times \frac{L}{4}$ square-like and $\frac{L}{4} \times L$ strip-like subsystems.
In this (sub)system configuration, the ratio of subsystem is fixed, and the scaling behavior with varying the total system size is investigated.
The ground state on square lattice exhibits a $S \sim L \log L$ scaling due to the presence of a codimension-one Fermi surface.
For the honeycomb lattice with Dirac cone, its entanglement entropy follows the area law as $S \sim L$.
For the Lieb lattice with a flat band, we observe a volume-law entanglement $S \sim L^2$.

Another considered (sub)system-cut configuration is to cut $L_{A} \times L$ strip-like subsystems in a size-fixed total system, with changing the ratio $u = L_A / L$.
The numerical results for different lattices with fixed total system size $120 \times 120$ are shown in Fig.~\ref{fig:GS_EE_fix}.
For square lattice, it behaves similarly with the $(1+1)$D critical system, with a $S \sim \log \sin (\pi u)$ entanglement scaling.
The ground state on honeycomb lattice gives a $S \sim 1/u$ scaling for $u < \frac{1}{2}$ and $S \sim 1/(1-u)$ for $u > \frac{1}{2}$.
For Lieb lattice, this (sub)system-cut configuration gives again a $S \sim u^2$ volume-law scaling.

\begin{figure}\centering
	\includegraphics[width=\columnwidth]{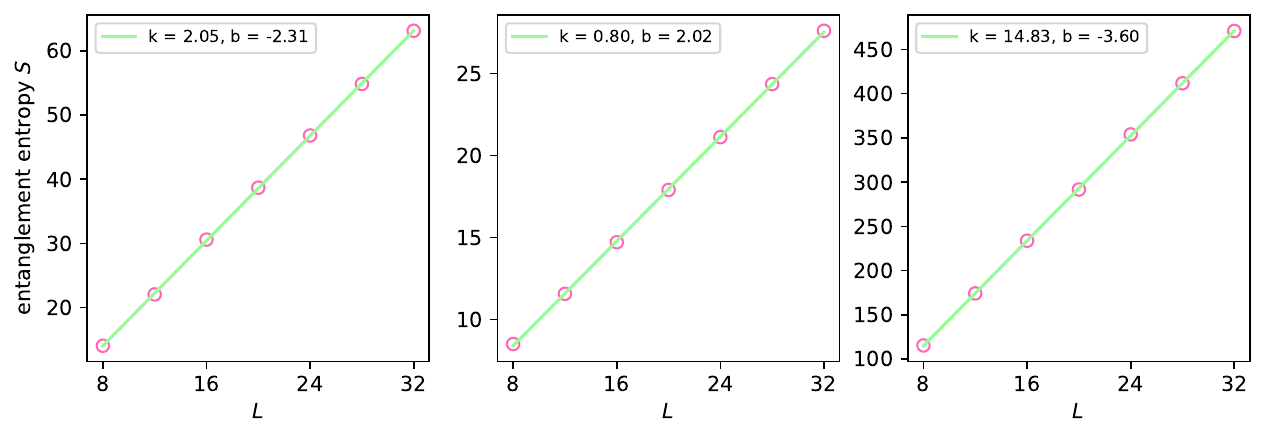}
	\caption{
		\label{fig:GS_thin_EE}
		The ground-state entanglement entropy of the homogeneous tight-binding model on (left) square, (middle) honeycomb, and (right) Lieb lattices at half-filling. 
		Here the total system size is set to be $300 \times L_y$ with changing $L_y$, and the subsystem is chosen to be half-cut with size $150 \times L_y$.
	}
\end{figure}

We can also consider the ``thin torus'' limit, i.e. the total system is quasi-1d with size $L_x \times L_y$ and $L_x \gg L_y$.
As shown in Fig.~\ref{fig:GS_thin_EE}, all three lattice geometries have a linear growing entanglement entropy $S \sim L_y$ under this (sub)system-cut configuration.

\subsection{Entanglement dynamics during unitary time evolution on different lattices}\label{app:unitary}

\begin{figure}\centering
	\includegraphics[width=0.85\columnwidth]{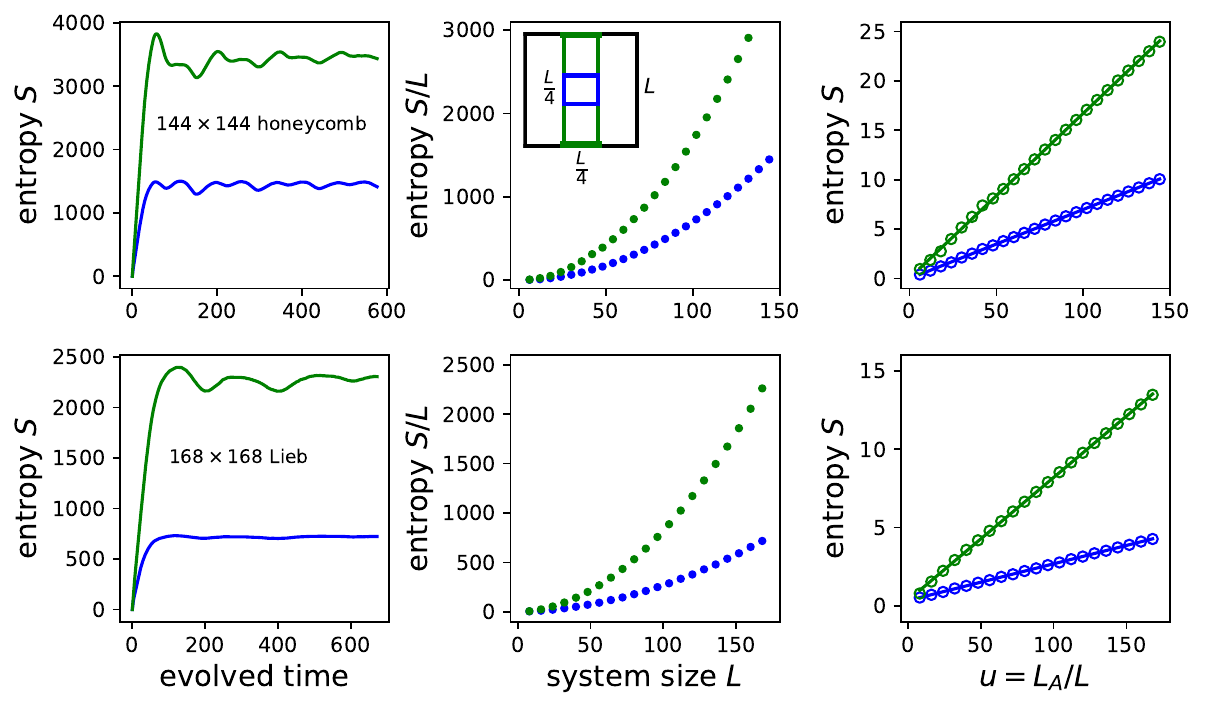}
	\caption{
		\label{fig:Scaling_RealEvoEE}
		Unitary dynamics on $L \times L$ (top) honeycomb and (bottom) Lieb lattices with half-filling. 
		As shown in the inset, we consider two kinds of subsystem configurations: $\frac{L}{4} \times \frac{L}{4}$ square-like subsystem (blue dots) and $\frac{L}{4} \times L$ strip-like subsystem (green dots). 
		(Left column) The time evolution of entanglement entropy $S(t)$.
		(Middle column) The entanglement entropy at large time $S(t\to \infty)$ plotted as a function of the system size $L$.
		(Right column) An alternative plot of $S(t\to \infty)/L$ as a function of $L$.
		The total system size is simulated upto $L = 144$ for honeycomb lattice and $L = 168$ for the Lieb lattice.
	}
\end{figure}

\begin{figure}\centering
	\includegraphics[width=0.6\columnwidth]{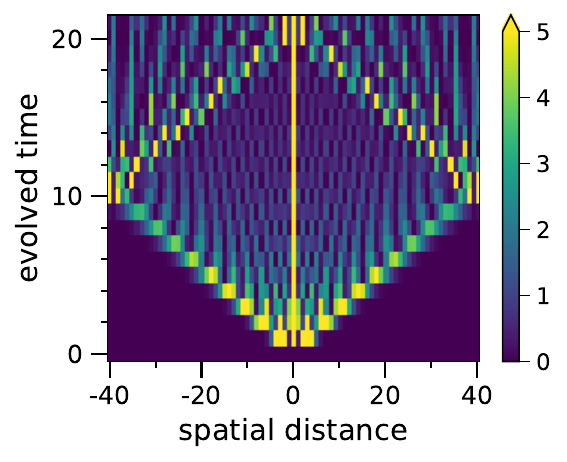}
	\caption{
		\label{fig:UnitEvo_mutinf_square}
		Spreading of mutual information $I_{A,B}$ during (homogeneous) unitary dynamics on the square lattice with half-filling. Here we consider subsystems $A$ and $B$ are two distant single columns, and the total system size is set to be $80 \times 80$.
	}
\end{figure}

We turn to discuss the time evolution of entanglement entropy during unitary dynamics on square, honeycomb, and Lieb lattices. 
For simplicity the initial state is chosen to have N\'eel-type order and no randomness is imposed into the quenching tight-binding Hamiltonian.
As shown in the left column of Fig.~\ref{fig:Scaling_RealEvoEE}, in all three cases (including the square lattice discussed in the main text, and the honeycomb and Lieb lattices presented here) the entanglement entropy grows rapidly to reach an asymptotically steady value.
Although there are some oscillations depending on the lattice details, at large time all three lattices exhibit universal volume-law entanglement.

As discussed in the main text, for (homogeneous) unitary time evolution of free fermions, the entanglement dynamics can be understood by the quasiparticle picture.
The entangled pairs propagate in the system to create mutual entanglement that can be measured by the mutual information $I_{A, B} = S_A + S_B - S_{A \cup B}$.
In Fig.~\ref{fig:UnitEvo_mutinf_square}, we present the numerical results of mutual information $I_{A, B}$ for two sub-regions $A$ and $B$ lives in a single column on the square lattice as a concrete example.
The mutual information at $t=0$ has zero values, since the system is initialized as a product state.
While applying the time evolution operator to the initial state,
the information is suddenly encoded locally.
As the time evolving, the mutual information exhibits a clear wave-front,
indicating the quasiparticles move with a fixed velocity.
After reaching the boundary, the quasiparticles reflect back and lead to the oscillations of entanglement entropy.
The observed phenomenon demonstrates that the unitary dynamics of entanglement develops by the quasiparticle propagation, akin to the $(1+1)$D integrable systems.

\section{Numerical results of the mixed nonunitary random dynamics with different settings}

\begin{figure}[b]\centering
	\includegraphics[width=0.85\columnwidth]{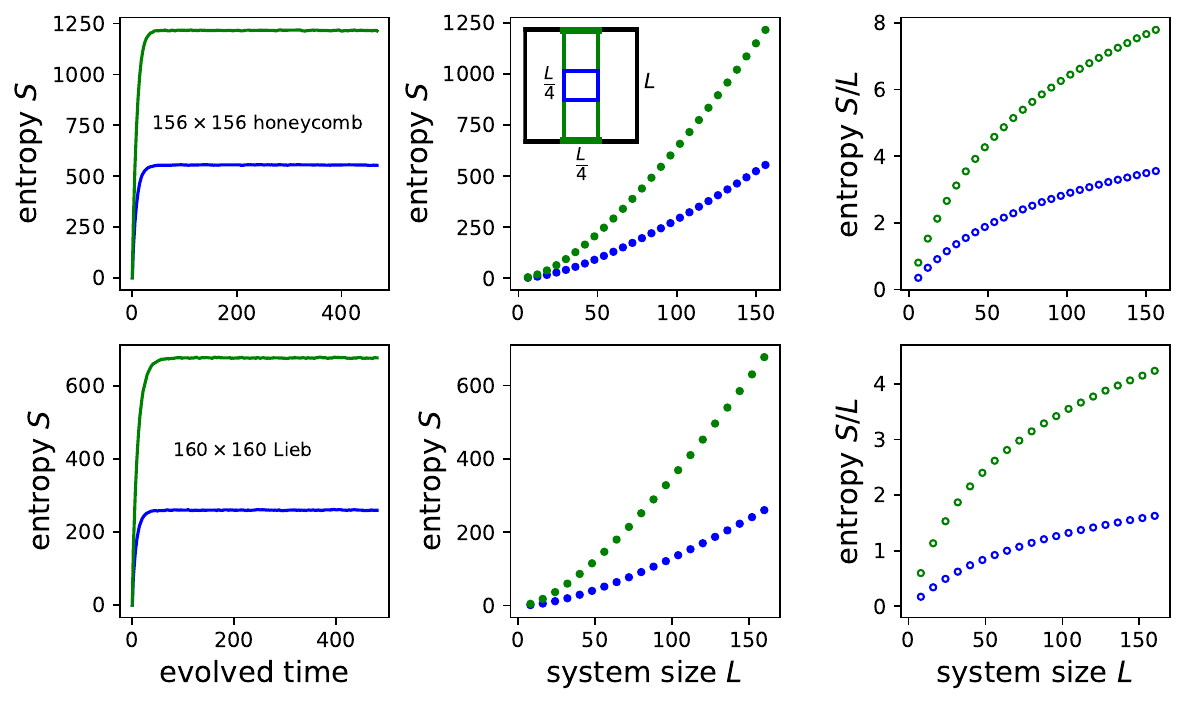}
	\caption{
		\label{fig:Scaling_NonUnitaryEE}
		Mixed nonunitary dynamics of entanglement entropy on $L \times L$ (top) honeycomb and (bottom) Lieb lattices with half-filling.
		As shown in the inset, we consider two kinds of subsystem configurations: $\frac{L}{4} \times \frac{L}{4}$ square-like subsystem (blue dots) and $\frac{L}{4} \times L$ strip-like subsystem (green dots). 
		(Left column) The time evolution of entanglement entropy $S(t)$.
		(Middle column) The entanglement entropy at large time $S(t\to \infty)$ plotted as a function of the system size $L$.
		(Right column) The entanglement entropy per length $S(t\to \infty)/L$ plotted as a function of $L$.
		Here we set the strength of imaginary random onsite potential $\beta=1$.
		The total system size is simulated up to $L = 156$ for honeycomb lattice and $L = 160$ for Lieb lattice.
	}
\end{figure}

\begin{figure}\centering
	\includegraphics[width=0.81\columnwidth]{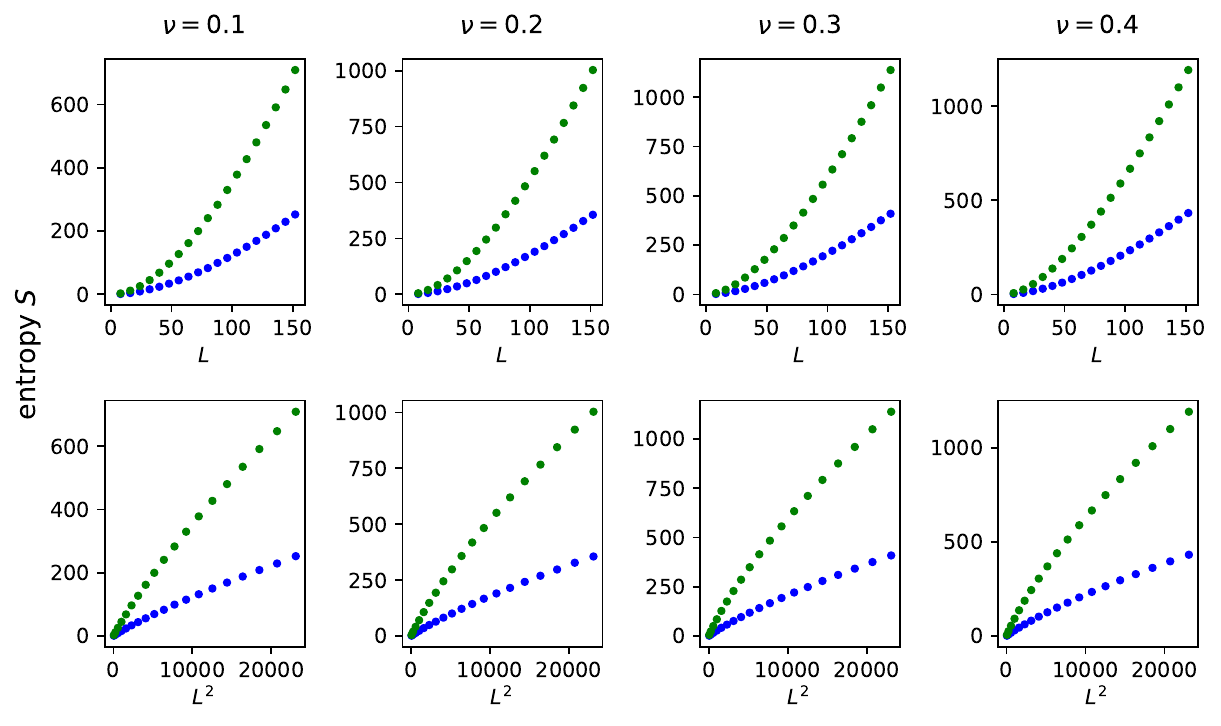}
	\caption{
		\label{fig:filling_Scaling_NonUnitaryEE_Square}
		Steady-state entanglement entropy of the nonunitary dynamics on $L \times L$ square lattice with different filling factor $\nu$, plotted as a function of the system size $L$.
		The subsystem configuration is similar to that in Fig.~\ref{fig:Scaling_NonUnitaryEE}.
		Here we set the strength of imaginary random onsite potential $\beta=1$.
	}
\end{figure}

\begin{figure}\centering
	\includegraphics[width=0.81\columnwidth]{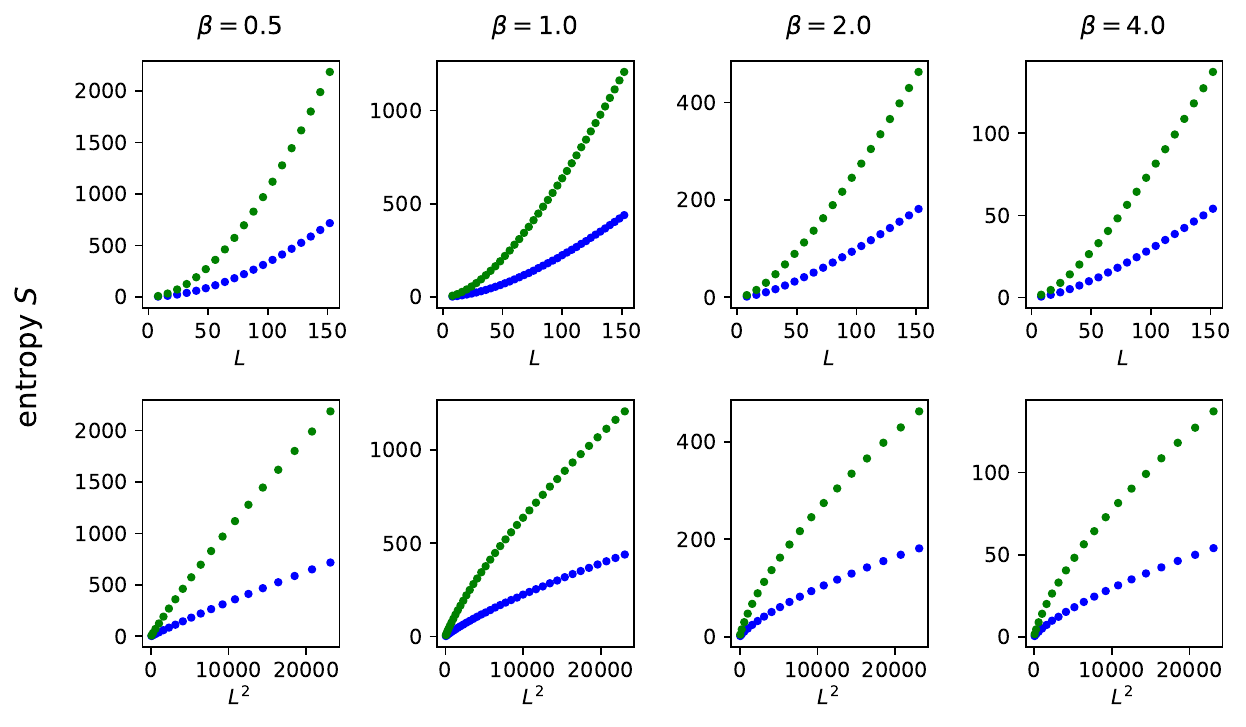}
	\caption{
		\label{fig:beta_Scaling_NonUnitaryEE_Square}
		Steady-state entanglement entropy of the nonunitary dynamics on $L \times L$ square lattice with different strength of imaginary onsite potential $\beta$,
		plotted as a function of (upper) the system size $L$ and (lower) the squared system size $L^2$.
		The subsystem configuration is similar to that in Fig.~\ref{fig:Scaling_NonUnitaryEE}.
		Here the system is chosen to be half-filled.
	}
\end{figure}

\begin{figure}\centering
	\includegraphics[width=0.6\columnwidth]{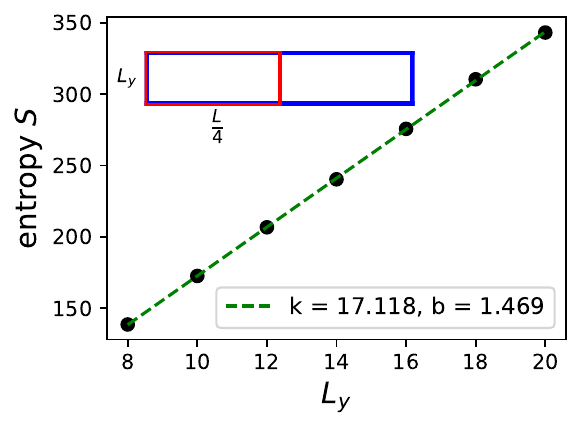}
	\caption{
		\label{fig:EE_steady_square_Lx1000LyChanged}
		The half-cut entanglement entropy of nonunitary steady state for $1000 \times L_y$ thin torus with changing $L_y$. The black dots represent the numerical data, and the green dashed line is the liner fit to $S(L_y) = kL_y + b$.
		Here the system is chosen to be half-filled, and the strength of imaginary random onsite potential $\beta$ is set to $1$.
	}
\end{figure}

In this appendix, we test the reported mixed nonunitary random dynamics in the main text with different settings of simulating parameters and lattice geometry.
It is found that the scaling behaviors of entanglement entropy, mutual information, and squared two-point correlation function are all robust under different settings.
This indicates that the emergent criticality in our designed model is quite universal.

\subsection{Entanglement entropy}\label{app:mixedEE}

\begin{figure}\centering
	\includegraphics[width=0.6\columnwidth]{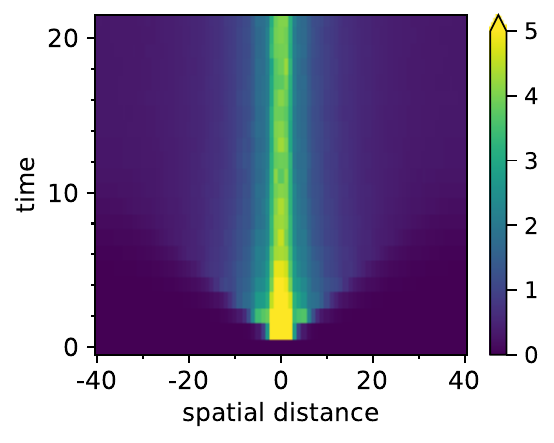}
	\caption{
		\label{fig:MixEvo_mutinf_square}
		Spreading of mutual information $I_{A,B}$ during the mixed nonunitary random dynamics on square lattice with half-filling. Here we consider subsystems $A$ and $B$ are two distant single columns, the total system size is chosen to be $80 \times 80$, and the strength of imaginary random onsite potential $\beta$ is set to be $1$.
	}
\end{figure}

\begin{figure}\centering
	\includegraphics[width=0.95\columnwidth]{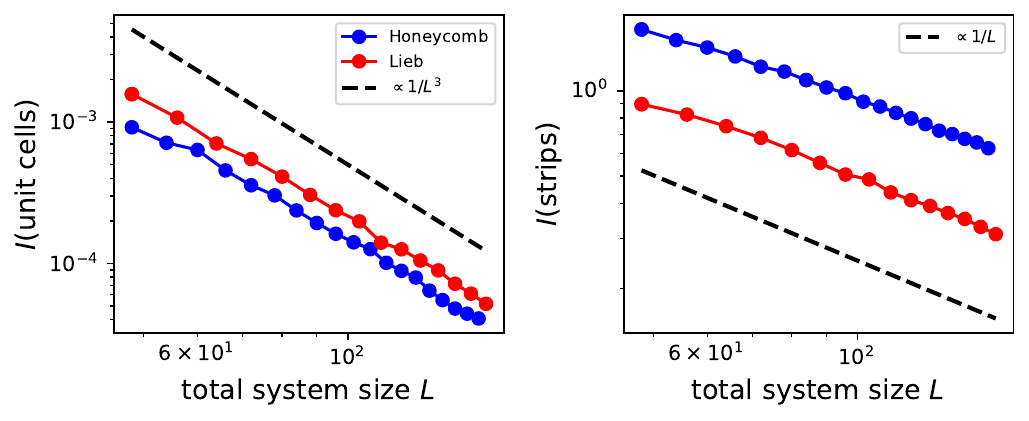}
	\caption{
		\label{fig:Mix_mutinf_other}
		The scaling of (left) mutual information between two unit cells $I_{\rm{unit \ cells}}$ and (right) mutual information between two strips $I_{\rm{strips}}$ for nonunitary steady state on honeycomb and Lieb lattices with half-filling.
		Here we set the strength of imaginary random onsite potential $\beta = 1$, and the distance between two subsystems is chosen to be $r = \frac{L}{2}$.
		The total system size is simulated up to $156 \times 156$ for honeycomb lattice and $160 \times 160$ for Lieb lattice.
	}
\end{figure}

For the mixed nonunitary dynamics, we have found that the steady state entanglement scaling form is robust for the choice of lattice geometry, filling factor $\nu$, and the strength of the imaginary onsite potential $\beta$.
At first, we present the numerical results of entanglement entropy for the steady state of the mixed nonunitary dynamics on different lattices in Fig.~\ref{fig:Scaling_NonUnitaryEE}.
Both honeycomb and Lieb lattices have very similar behavior of the entanglement entropy with the square lattice reported in the main text.
Second, we calculate the mixed nonunitary dynamics on the square lattice with different filling factors.
The results are plotted in Fig.~\ref{fig:filling_Scaling_NonUnitaryEE_Square}.
It is clear to see that the steady-state entanglement entropy has very similar behavior for different filling.
For $\nu \ge 0.2$, even the absolute value of entropy is close for different values of filling factor.
Third, we investigate the influence of the strength of imaginary onsite potential $\beta$.
As shown in Fig.~\ref{fig:beta_Scaling_NonUnitaryEE_Square}, we find that the different values of $\beta$ lead to universal steady-state entanglement scaling.
Moreover, we consider the thin torus limit of the square lattice with $L_x \gg L_y$.
As shown in Fig.~\ref{fig:EE_steady_square_Lx1000LyChanged},
the steady-state entanglement entropy under thin torus limit scales as $S \sim L_y$, akin to the ground-state case.

\subsection{Mutual information in the steady state}\label{app:mixedMI}

We turn to investigate mutual information during mixed nonunitary dynamics.
In Fig.~\ref{fig:MixEvo_mutinf_square}, we plot the time evolution of mutual information on the square lattice.
The unitary background still plays a role of the population of mutual information, and a weak signature of a linear dispersion of the entanglement propagation can be observed.
However, due to the nonunitary operations, the plane-wave-like behavior of the mutual information is destroyed.
Instead, it spreads in the spatial space very weakly (but fast), and no reflection is located at the boundaries.
As a result, different from the unitary case, the nonunitary dynamics flows to a steady state without oscillations of entanglement entropy in the time domain.

We further consider different lattice geometries, including honeycomb and Lieb lattices.
In Fig.~\ref{fig:Mix_mutinf_other}, we show the scaling of mutual information for the steady state.
Instead of the two-point correlation, here we consider the mutual information between two unit cells $I({\rm{unit \ cells}})$, which contains nearest two sites for the honeycomb lattice and three for the Lieb lattice.
For mutual information between two strips $I({\rm{strips}})$, we choose the subsystem size to be $2 \times L$ to make sure that the unit cells are not cut apart.
We find that both the honeycomb and Lieb lattices give the same scaling behavior of mutual information as the square lattice, with $I({\rm unit \ cells}) \sim 1/L^3$ and $I({\rm strips}) \sim 1/L$.
This supports that the underlying quasiparticle picture is universal for different lattice geometries.

\begin{figure}\centering
	\includegraphics[width=0.95\columnwidth]{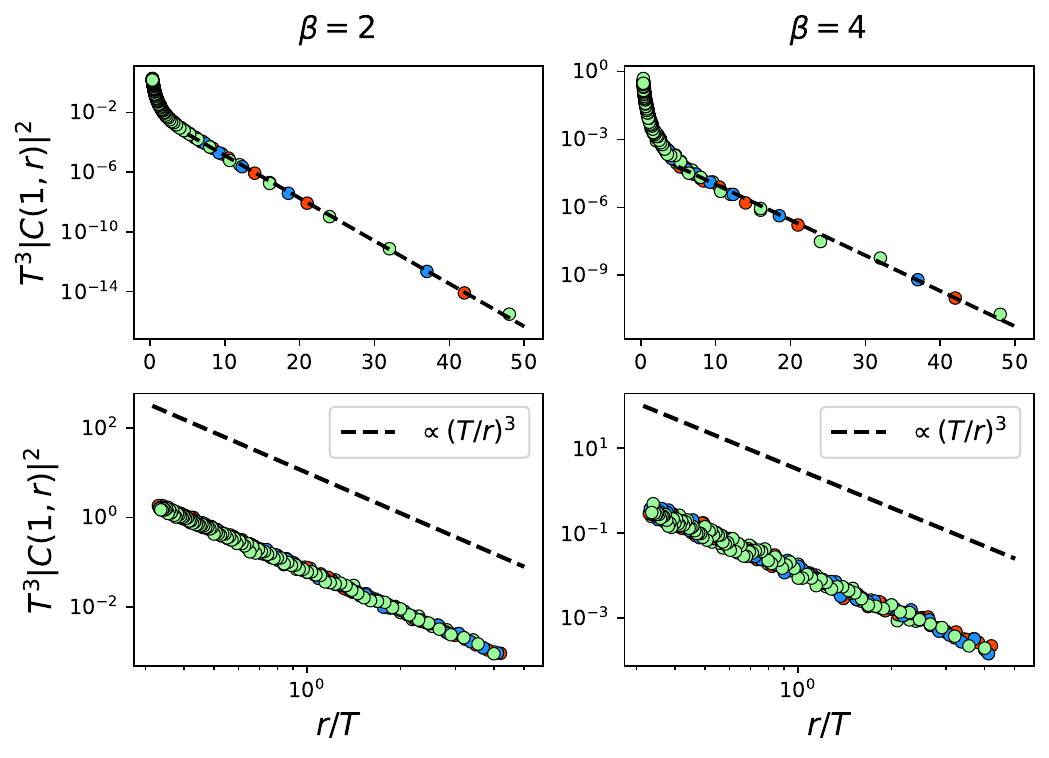}
	\caption{
		\label{fig:MixCorr_square_beta}
		Square two-point correlation function $|C(1,r)|^2$ during the mixed nonunitary random dynamics on $L \times L$ square lattice with (left column) $\beta = 2$ and (right column) $\beta = 4$.
		(Upper) The early time evolution of $T^2|C(1,r)|^2$ at the limit $T \ll r$ is plotted in semi-log scale.
		(Lower) Later (before reaching steadiness) dynamics of $T^2|C(1,r)|^2$ is plotted in log-log scale.
		Here we consider half-filling.
		The total system size is set to be $L=80$, $96$, and $128$, data calculated with different $L$ is represented by different colors. 
	}
\end{figure}

\subsection{Dynamics of two-point correlation functions}\label{app:mixedCorr}

In Fig.~\ref{fig:MixCorr_square_beta}, we plot the dynamics of two-point correlation functions during mixed nonunitary random dynamics on square lattice with various values of $\beta$.
We find that the scaling behavior
\begin{equation}
|C(r,T)|^2 \sim 
\begin{cases}
{T^{-3}\exp(-c r/T)}, &\qquad T \ll r  \\
1 / r^{-3}, &\qquad T \gg r
\end{cases}
\end{equation}
is robust under different settings.
The insensitivity of the strength of imaginary random onsite potential supports the argument that the diffusive part (comes from the unitary dynamics) is not important for the spreading of correlation function.

\begin{figure}\centering
	\includegraphics[width=\columnwidth]{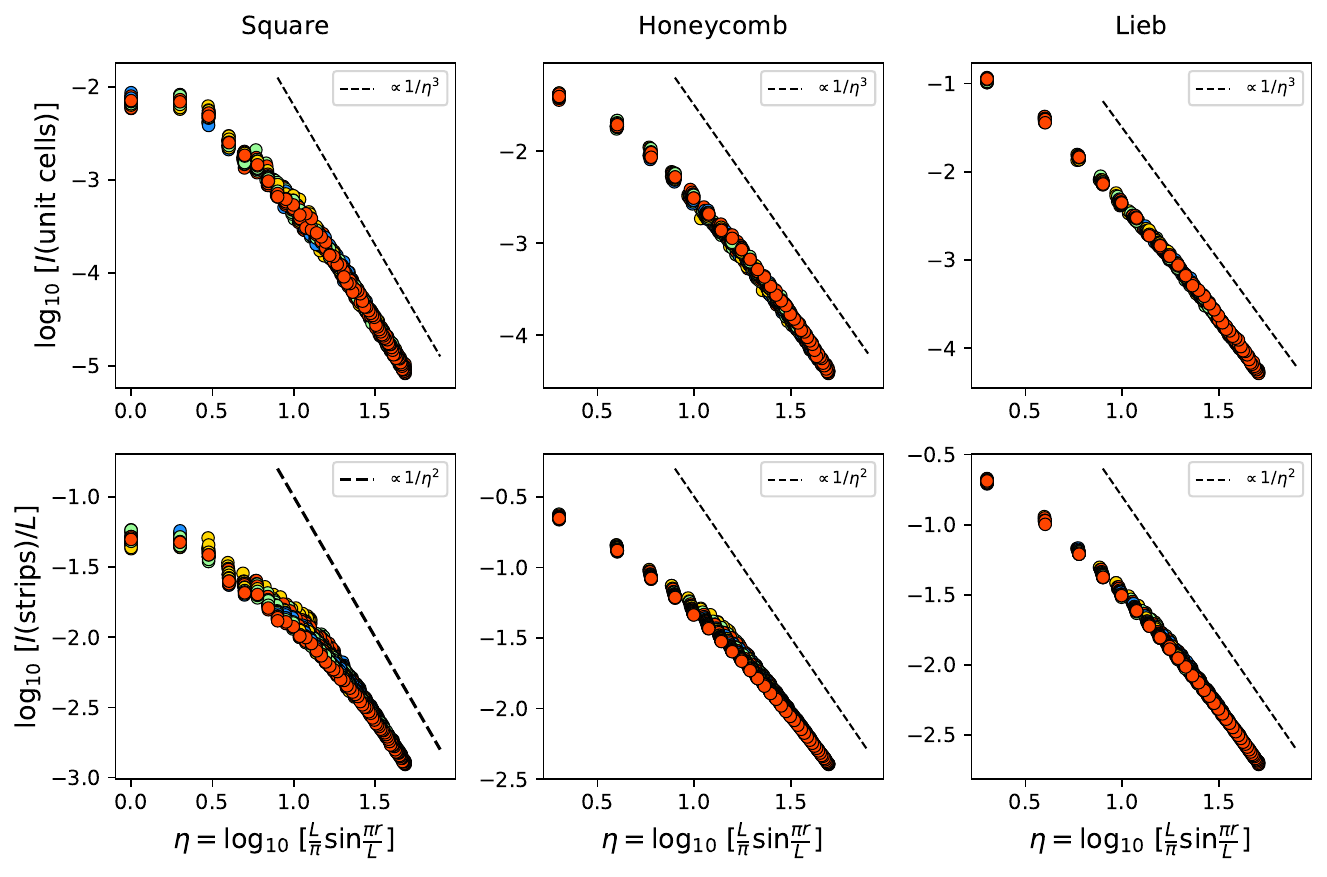}
	\caption{
		\label{fig:DataCollapse_Mix_RandOnsite}
		The data collapse within the scaling form in Eq.~\eqref{eq:C_scaling} for (upper) mutual information between two unit cells $I({\rm{unit \ cells}})$ and (lower) between two strips $I({\rm{strips}})$ for \textbf{mixed nonunitary} steady state on (left) square, (middle) honeycomb and (right) Lieb lattices.
		The total system size is simulated upto $176 \times 176$, $156 \times 156$ and $160 \times 160$, respectively.
		Here the system is chosen to be half-filled, and the randomness strength is set to be $\beta=1$.
	}
\end{figure}

\begin{figure}\centering
	\includegraphics[width=\columnwidth]{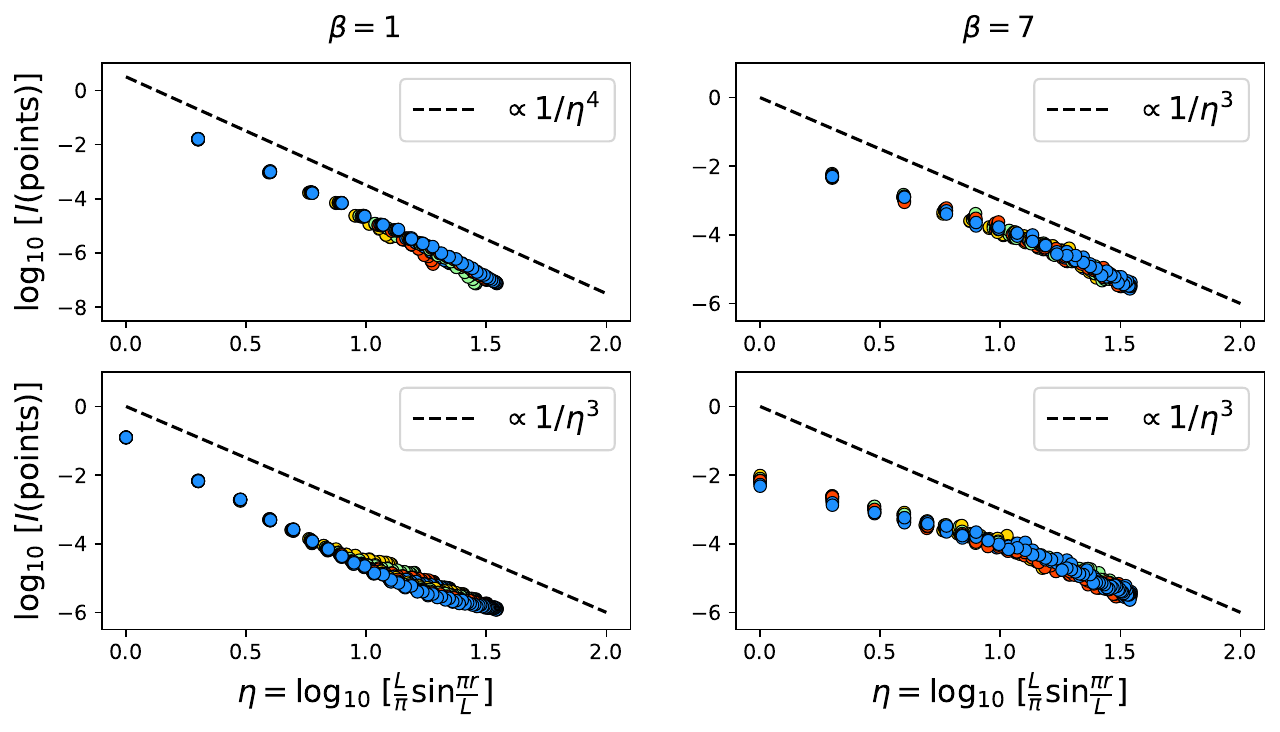}
	\caption{
		\label{fig:DataCollapse_PureImag_RandOnsite}
		The data collapse within the scaling form in Eq.~\ref{eq:C_scaling} for  information between two points and for \textbf{pure imaginary} steady state on (upper) square and (bottom) honeycomb lattices with half-filling.
		Here the total system size is simulated upto $120 \times 120$.
	}
\end{figure}

\section{Possible finite-size scaling form of the quantum correlations in the nonunitary steady state}

In this appendix, we propose a possible finite-size scaling form of quantum correlations in the steady state of the nonunitary random dynamics of $(2+1)$D free fermions.
The basic idea is to extend the known results of $(1+1)$D case into higher dimension.
In $(1+1)$D, it has been found that the mixed nonunitary dynamics is described by CFT, where the two-point correlation functions have the exact form
\begin{equation}\label{eq:C_scaling}
|C(1,r)|^2 \equiv |\langle c^\dagger_1 c_r \rangle|^2 \propto [L \sin(\pi r/L)]^{-a} .
\end{equation}

Although the exact form in the $(2+1)$D case is unknown (even the existence of the conformal symmetry cannot be confirmed), there are still notable facts that lead to a direct extension of the scaling form obtained in $(1+1)$D:

1. Numerically we find that the entanglement entropy of a $L_A \times A$ strip-like subsystem in a $L \times L$ finite periodic system scales as $S \propto \ln \sin \frac{\pi L_A}{L}$, akin to the result of cutting a single interval in $(1+1)$D.

2. The mutual information between two narrow strips in $(2+1)$D scales as $I({\rm{strips}}) \propto 1/r^2$, similar to the squared two-point correlation function in $(1+1)$D.

3. The form in Eq.~\ref{eq:C_scaling} satisfies the boundary condition and gives expected power-law scaling $I(r) \propto r^{-a}$ at the limit $r \ll L$.

Thus, it is reasonable to consider the mutual information between two strips in $(2+1)$D can have the same finite-size scaling form with the squared two-point correlation function in $(1+1)$D.
We have tested this conjectured form on different lattice geometries, and the results are shown in Figs.~\ref{fig:DataCollapse_Mix_RandOnsite} and~\ref{fig:DataCollapse_PureImag_RandOnsite}.
Surprisingly, we find that the scaling form not only works for the mutual information between two strips, but also for two points.
Both $I({\rm{points}})$ and $I({\rm{strips}})$ collapse into single curve with the slope close to their critical exponent.
Based on these, we argue that Eq.~\eqref{eq:C_scaling} could be the correct finite-size scaling form for quantum correlations in the steady state of the nonunitary random dynamics.

\begin{figure}\centering
	\includegraphics[width=0.95\columnwidth]{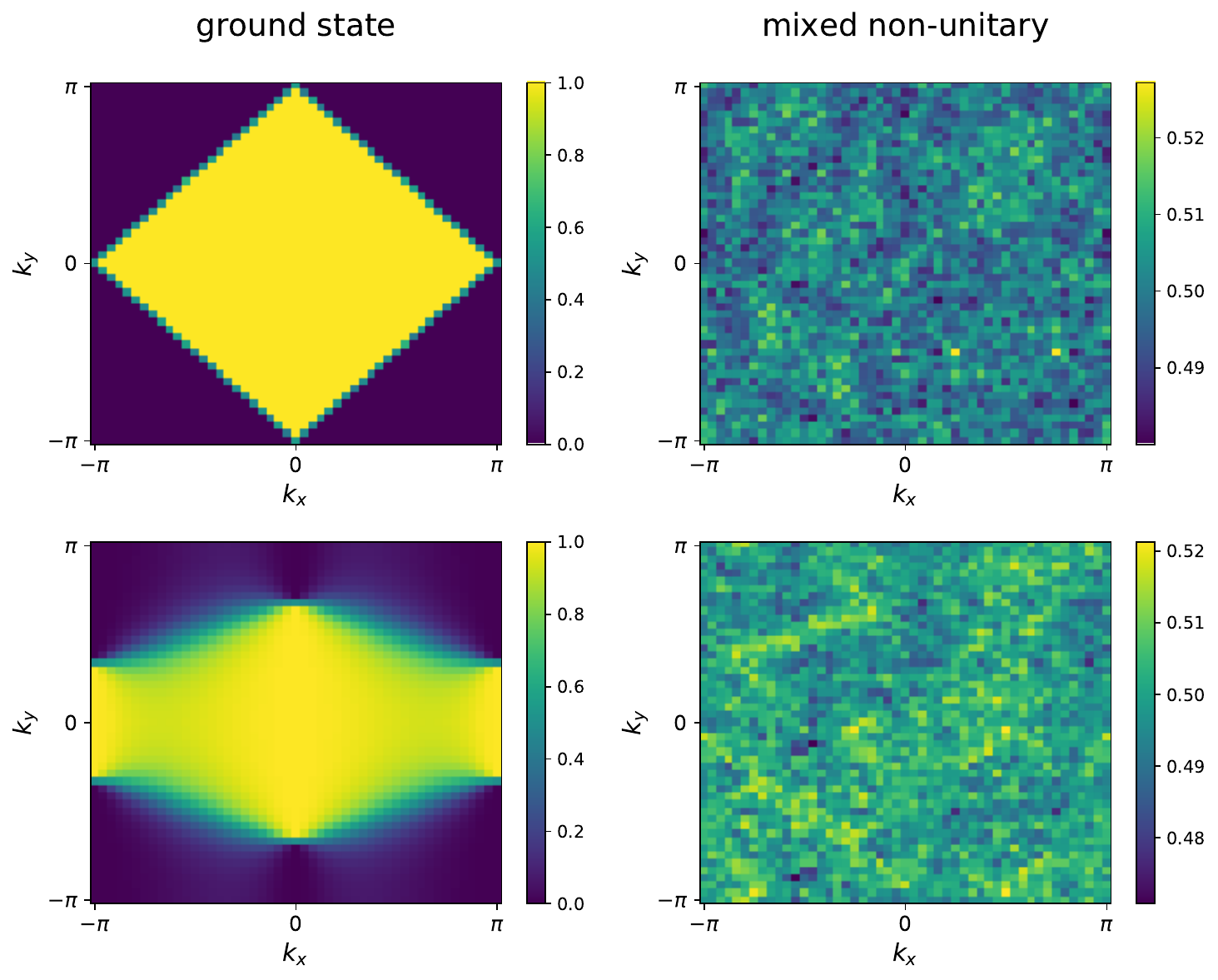}
	\caption{
		\label{fig:MixEvo_nk}
		Occupation number in momentum space $n(k)$ for (left column) \textbf{ground state} and (right column) steady state of the \textbf{mixed nonunitary} random dynamics on $48 \times 48$ (top) square and (bottom) honeycomb lattices with half-filling.
	}
\end{figure}

\begin{figure}\centering
	\includegraphics[width=0.95\columnwidth]{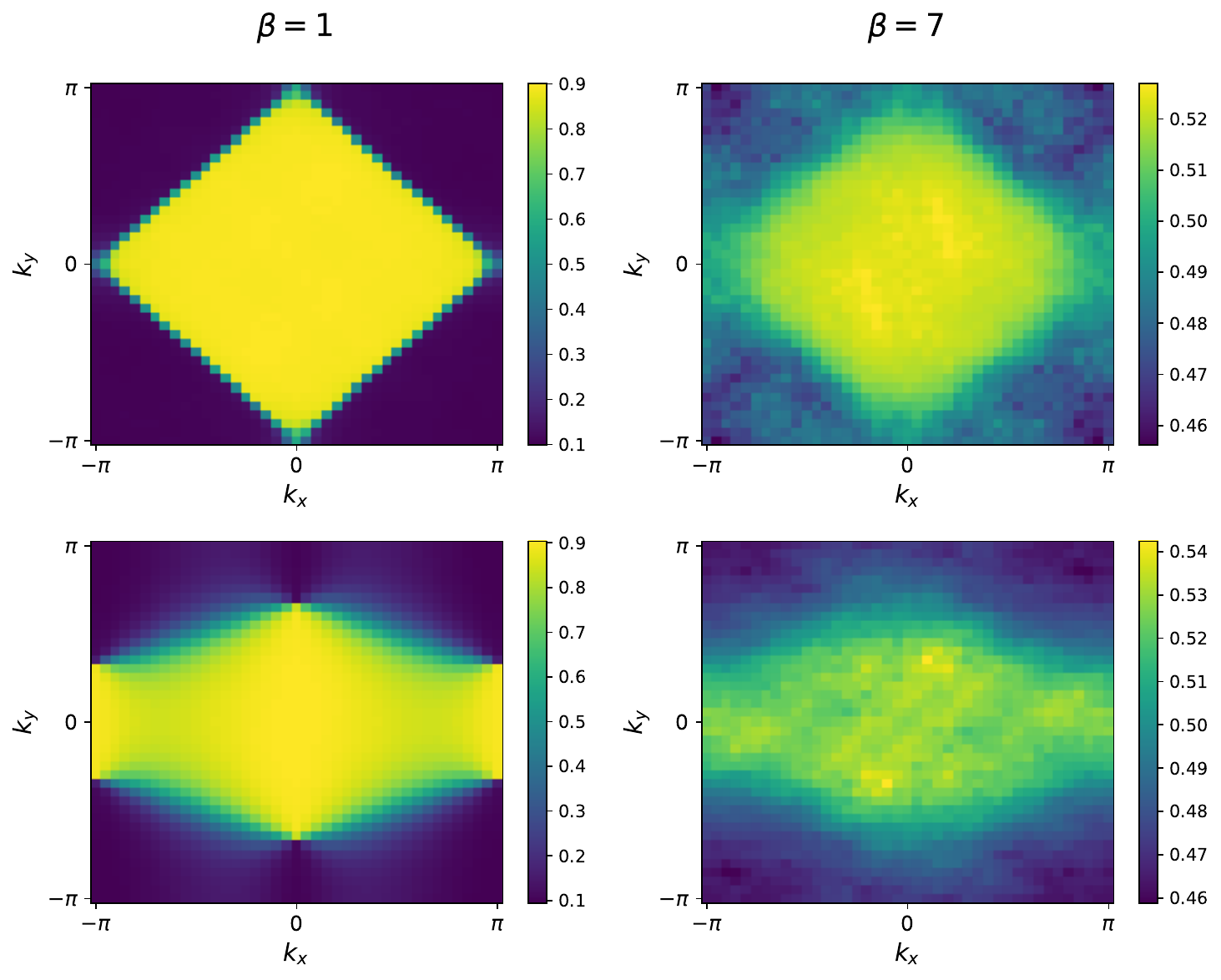}
	\caption{
		\label{fig:PureImag_TBRand_nk}
		Occupation number in momentum space $n(k)$ for steady state of the \textbf{pure imaginary} random dynamics on $40 \times 40$ (top) square and (bottom) honeycomb lattices with half-filling.
	}
\end{figure}

\section{Absence of a Fermi surface in the dynamic steady state}\label{app:absence}

In this work, we find that nonunitary random dynamics of $(2+1)$D free fermions flows to a dynamic steady state with entanglement entropy $S \sim L\log L$ and mutual information $I({\rm points}) \sim 1/r^3$. 
These scaling behaviors are quite similar with the ground state of Fermi liquids.
However, there is no concept of Fermi surface in our model of random dynamics.
This is important because it will indicate a very different mechanism of the formation of logarithmic entanglement entropy during nonunitary random dynamics.
In this appendix, we present direct numerical evidence that a Fermi surface is absent in the designed dynamic steady state.
In particular, we consider the occupation number in momentum space 
\begin{equation}
n(k) = \sum_{a, b} \exp[-ik_x(x_a-x_b)] \exp[-ik_y(y_a-y_b)] \langle c^\dagger_a c_b \rangle ,
\end{equation}
as an indicator of the absence or presence of a Fermi surface.
As shown in Fig.~\ref{fig:MixEvo_nk}, the occupation $n(k)$ in the steady state of mixed nonunitary random dynamics is totally disordered, and has values close to $0.5$ for all momentum $(k_x,k_y)$.
This is a strong evidence that in our model a Fermi surface is absent.
Different for the randomness in mixed nonunitary dynamics, for pure imaginary dynamics, we observe a similar pattern as the ground state (see Fig.~\ref{fig:PureImag_TBRand_nk}). This could be caused by the imaginary background dynamics, which flows to the ground state with a finite Fermi surface. However, we notice that both protocols have the same entanglement scaling. 
This indicates that the pattern of $n(k)$ is not important, and only the random imaginary onsite potential is responsible for the emergent nontrivial entanglement structure.


\begin{widetext}

	\section{Nonlinear master equation of two-point correlation function in $2+1$ dimensions}\label{app:master}
	In this appendix, we introduce a simple nonlinear master equation~\cite{chen2020emergent} that can effectively describe the mixed nonunitary dynamics reported in the main text.
	We find that the Brownian dynamics in $(2+1)$D leads to 
	\begin{equation}\label{eq:CorrScaling}
	|C(1,r)|^2 \sim
	\begin{cases}
	t^{-3}\exp(-cr/t), &\qquad t \ll r \\
	1/r^3, &\qquad t \gg r ,
	\end{cases}
	\end{equation}which is consistent with the numerical results of the mixed nonunitary dynamics.
	Based on this, we conclude that the nonlinear master equation could be used to describe the $(2+1)$D mixed nonunitary dynamics.
	
	Here we consider the dynamics of two-point correlation function as a nonunitary Brownian dynamics which contains unitary and imaginary time evolution.
	We want to calculate 
	\begin{equation}
	\frac{d|C_{a, b}|^2}{dt} = \frac{d C_{a, b} d C_{a, b} + \frac{1}{2} \left( d^2 C_{a, b} C_{a, b} + C_{a, b} d^2 C_{a, b} \right) }{dt} .
	\end{equation}

	We first write the equation of motion for the unitary dynamics, where we have 
	\begin{equation}
	dC = i\left[dH^{\rm{real}}, C \right] \qquad \Longrightarrow \qquad 
	dC_{a, b}  =  i \sum_m \left\{ dH^{\rm{real}}_{a, m} C_{m, b} - C_{a, m} dH^{\rm{real}}_{m, b} \right\}
	\end{equation}
	and
	\begin{equation}
	\begin{aligned}
	& \qquad \qquad \qquad \qquad {d^2C} = - \left[ dH^{\rm{real}}, \left[dH^{\rm{real}}, C \right] \right]  \\ \Longrightarrow 
	{d^2C_{a, b}}  =  - \sum_{m,n} &
	\left\{ 
	dH^{\rm{real}}_{a, m} dH^{\rm{real}}_{m, n} C_{n, b}
	- dH^{\rm{real}}_{a, m} C_{m, n} dH^{\rm{real}}_{n, b}
	- dH^{\rm{real}}_{a, n} C_{n, m} dH^{\rm{real}}_{m, b}
	+ C_{a, n} dH^{\rm{real}}_{n, m} dH^{\rm{real}}_{m, b} 
	\right\}
	\end{aligned}
	\end{equation}
	with
	\begin{equation}
	dH^{\rm{real}} = \sum_{x,y} (c_{x+1,y}^\dagger c_{x,y} dW_{x,y}^1 + c_{x,y}^\dagger c_{x+1,y} d{\overline{W}}_{x,y}^1 + c_{x,y+1}^\dagger c_{x,y} dW_{x,y}^2 + c_{x,y}^\dagger c_{x,y+1} d{\overline{W}}_{x,y}^2 ) ,
	\end{equation}
	where $\{dW_{x,y}^1, d{\overline{W}}_{x,y}^1 \}$ and $\{dW_{x,y}^2, d{\overline{W}}_{x,y}^2 \}$ are two independent sets of the Brownian motion in $x$ and $y$ directions, respectively.
	We can set that the two independent stochastic variables $dW^1$ and $dW^2$ have the same strength of randomness $A$, i.e.,
	\begin{equation}
	dW_{a}^1 d{\overline{W}}_{b}^1 = dW_{a}^2 d{\overline{W}}_{b}^2 = A dt \delta_{a,b} \qquad {\rm{and}} \qquad dW_{a}^1 d{\overline{W}}_{b}^2 = 0 .
	\end{equation}

	For the first derivative, we have
	\begin{equation}
	\begin{aligned}
	\sum_m dH^{\rm{real}}_{a, m} C_{m, b} & = 
	[c_a^\dagger c_{a-\delta x} ] dW^1_{a-\delta x} C_{a-\delta x, b}
	+ [c_a^\dagger c_{a+\delta x} ] d{\overline{W}}^1_{a} C_{a+\delta x, b}
	+ [c_a^\dagger c_{a-\delta y} ] dW^2_{a-\delta y} C_{a-\delta y, b}
	+ [c_a^\dagger c_{a+\delta y} ] d{\overline{W}}^2_{a} C_{a+\delta y, b} ,
	\\
	\sum_m C_{a, m} dH^{\rm{real}}_{m, b} & = 
	[c_{b+\delta x}^\dagger c_{b} ] dW^1_{b} C_{a, b+\delta x}
	+ [c_{b-\delta x}^\dagger c_{b} ] d{\overline{W}}^1_{b-\delta x} C_{a, b-\delta x}
	+ [c_{b+\delta y}^\dagger c_{b} ] dW^2_{b} C_{a, b+\delta y}
	+ [c_{b-\delta y}^\dagger c_{b} ] d{\overline{W}}^2_{b-\delta y} C_{a, b-\delta y} .
	\end{aligned}
	\end{equation}
	From this we can calculate a single matrix element of $C_{a, b}$ as
	\begin{equation}
	\begin{aligned}
	dC_{a, b}
	= & i \sum_m \left\{ dH^{\rm{real}}_{a, m} C_{m, b} - C_{a, m} dH^{\rm{real}}_{m, b} \right\} \\
	= & i (
	dW^1_{a-\delta x} C_{a-\delta x, b} + d{\overline{W}}^1_{a} C_{a+\delta x, b}
	+ dW^2_{a-\delta y} C_{a-\delta y, b} + d{\overline{W}}^2_{a} C_{a+\delta y, b} \\
	& \quad - dW^1_{b} C_{a, b+\delta x} - d{\overline{W}}^1_{b-\delta x} C_{a, b-\delta x}
	- dW^2_{b} C_{a, b+\delta y} - d{\overline{W}}^2_{b-\delta y} C_{a, b-\delta y}
	) .
	\end{aligned}
	\end{equation}
	Then,
	\begin{equation}
	\begin{aligned}
	{d C_{b,a} d C_{a,b}} = & 
	- (
	dW^1_{a-\delta x} C_{a-\delta x, b} + d{\overline{W}}^1_{a} C_{a+\delta x, b}
	+ dW^2_{a-\delta y} C_{a-\delta y, b} + d{\overline{W}}^2_{a} C_{a+\delta y, b} \\
	& \qquad - dW^1_{b} C_{a, b+\delta x} - d{\overline{W}}^1_{b-\delta x} C_{a, b-\delta x}
	- dW^2_{b} C_{a, b+\delta y} - d{\overline{W}}^2_{b-\delta y} C_{a, b-\delta y}
	) \\
	& \quad \times ( 
	dW^1_{b-\delta x} C_{b-\delta x, a} + d{\overline{W}}^1_{b} C_{b+\delta x, a}
	+ dW^2_{b-\delta y} C_{b-\delta y, a} + d{\overline{W}}^2_{b} C_{b+\delta y, a} \\
	& \qquad - dW^1_{a} C_{b, a+\delta x} - d{\overline{W}}^1_{a-\delta x} C_{b, a-\delta x}
	- dW^2_{a} C_{b, a+\delta y} - d{\overline{W}}^2_{a-\delta y} C_{b, a-\delta y} 
	)
	\\
	= & {Adt} (
	|C_{a-\delta x, b}|^2 + |C_{a+\delta x, b}|^2 + |C_{a-\delta y, b}|^2 + |C_{a+\delta y, b}|^2
	+ |C_{a, b-\delta x}|^2 + |C_{a, b+\delta x}|^2 + |C_{a, b-\delta y}|^2 + |C_{a, b+\delta y}|^2 \\
	& \qquad - 2 C_{b, b} C_{b+\delta x, b+ \delta x} \delta_{a-\delta x,b}
	- 2 C_{b, b} C_{b-\delta x, b- \delta x} \delta_{a+\delta x,b}
	- 2 C_{b, b} C_{b+\delta y, b+ \delta y} \delta_{a-\delta y,b}
	- 2 C_{b, b} C_{b-\delta y, b- \delta y} \delta_{a+\delta y,b}
	) .
	\end{aligned}
	\end{equation}
	
	For second derivative, we have
	\begin{equation}
	\begin{aligned}
	\sum_{m, n} dH_{a,m} dH_{m,n} C_{n,b} =  \sum_{m, n} dH_{a,n} dH_{n,m} C_{m,b}
	=  \ &
	dH_{a,a-\delta x} dH_{a-\delta x,a} C_{a,b}
	+ dH_{a,a+\delta x} dH_{a+\delta x,a} C_{a,b} \\
	& + dH_{a,a-\delta y} dH_{a-\delta y,a} C_{a,b}
	+ dH_{a,a+\delta y} dH_{a+\delta y,a} C_{a,b} \\
	= \ &
	4 C_{a,b} , \\
	\\
	\sum_{m, n} dH_{a,m} C_{m,n} dH_{n,b} =  \sum_{m, n} dH_{a,m} C_{m,n} dH_{n,b}
	= \ &
	dH_{a,a-\delta x} C_{a-\delta x,b-\delta x} dH_{b-\delta x,b}
	+ dH_{a,a+\delta x} C_{a+\delta x,b+\delta x} dH_{b+\delta x,b} \\
	& + dH_{a,a-\delta y} C_{a-\delta y,b-\delta y} dH_{b-\delta y,b}
	+ dH_{a,a+\delta y} C_{a+\delta y,b+\delta y} dH_{b+\delta y,b} \\
	= \ & \delta_{a,b} (C_{b-\delta x, b-\delta x} + C_{b+\delta x, b+\delta x} + C_{b-\delta y, b-\delta y} + C_{b+\delta y, b+\delta y}) ,
	\end{aligned}
	\end{equation}
	where we have used the orthogonality of the stochastic variables $dW^1$ and $dW^2$.
	Then, we have
	\begin{equation}
	\frac{d^2C_{a,b}}{Adt}  = 8 C_{a,b} - 2\delta_{a,b} (C_{b-\delta x, b-\delta x} + C_{b+\delta x, b+\delta x} + C_{b-\delta y, b-\delta y} + C_{b+\delta y, b+\delta y}) 
	\end{equation}
	which leads to
	\begin{equation}
	\frac{d^2C_{b,a} C_{a,b} + C_{b,a} d^2 C_{a,b}}{2Adt} = 8 |C_{a,b}|^2 - 2 \delta_{a,b} 
	(C_{b-\delta x, b-\delta x}C_{b,b} + C_{b+\delta x, b+\delta x}C_{b,b} + C_{b-\delta y, b-\delta y}C_{b,b} + C_{b+\delta y, b+\delta y}C_{b,b}) .
	\end{equation}
	
	Finally, we have
	\begin{equation}\label{eq:UnitME2d}
	\begin{aligned}
	\frac{d|C_{a,b}|^2}{Adt} = & \ 
	|C_{a-\delta x, b}|^2 + |C_{a+\delta x, b}|^2 + |C_{a-\delta y, b}|^2 + |C_{a+\delta y, b}|^2
	+ |C_{a, b-\delta x}|^2 + |C_{a, b+\delta x}|^2 + |C_{a, b-\delta y}|^2 + |C_{a, b+\delta y}|^2 \\
	& \ - 2 C_{b, b} C_{b+\delta x, b+ \delta x} \delta_{a-\delta x,b}
	- 2 C_{b, b} C_{b-\delta x, b- \delta x} \delta_{a+\delta x,b}
	- 2 C_{b, b} C_{b+\delta y, b+ \delta y} \delta_{a-\delta y,b}
	- 2 C_{b, b} C_{b-\delta y, b- \delta y} \delta_{a+\delta y,b} \\
	& \ + 8 |C_{a,b}|^2 - 2 \delta_{a,b} 
	(C_{b-\delta x, b-\delta x}C_{b,b} + C_{b+\delta x, b+\delta x}C_{b,b} + C_{b-\delta y, b-\delta y}C_{b,b} + C_{b+\delta y, b+\delta y}C_{b,b}) .
	\end{aligned}
	\end{equation}
	It is easy to check that the sum of squared two-point correlation function is 0, and the trace of the matrix (which is the total particle number) is conserved in the dynamics.
	
	
	For imaginary dynamics, we have
	\begin{equation}
	dC = -\{dH^{\rm{imag}}, C\} + 2CdH^{\rm{imag}}C
	\Longrightarrow 
	{dC_{a,b}} = - 
	\sum_m \left\{ dH^{\rm{imag}}_{a, m} C_{m, b}
	+ C_{a, m} dH^{\rm{imag}}_{m, b} \right\} 
	+ 2 \sum_{m,n} C_{a,m} dH^{\rm{imag}}_{m, n} C_{n, b}
	\end{equation}
	and
	\begin{equation}
	\begin{aligned}
	d^2C = & \ -dH^{\rm{imag}} (-\{dH^{\rm{imag}}, C\} + 2CdH^{\rm{imag}}C) 
	- (-\{dH^{\rm{imag}}, C\} + 2CdH^{\rm{imag}}C) dH^{\rm{imag}} \\
	& + 2 (-\{dH^{\rm{imag}}, C\} + 2CdH^{\rm{imag}}C) C dH^{\rm{imag}}
	+ 2 C dH^{\rm{imag}} (-\{dH^{\rm{imag}}, C\} + 2CdH^{\rm{imag}}C) \\
	= & \ -dH^{\rm{imag}} dC - dC dH^{\rm{imag}} 
	+ 2dC dH^{\rm{imag}} C +2 C dH^{\rm{imag}} dC \\
	\Longrightarrow \qquad d^2C_{a,b} = & \ 
	- \sum_{m} (
	dH^{\rm{imag}}_{a,m} dC_{m,b}
	+ dC_{a,m}dH^{\rm{imag}}_{m,b} 
	)
	+ 2 \sum_{m,n} (
	dC_{a,m} dH^{\rm{imag}}_{m,n} C_{n,b}
	+ C_{a,m}dH^{\rm{imag}}_{m,n} dC_{n,b}
	) ,
	\end{aligned}
	\end{equation}
	where
	\begin{equation}
	dH^{\rm{imag}} = \sum_{x,y} c_{x,y}^{\dagger} c_{x,y} dW^3_{x,y} , \qquad {\rm{with}} \qquad
	dW_a^3 dW_b^3 = \delta_{a,b} B dt .
	\end{equation}
	
	From this we have
	\begin{equation}
	dC_{a,b} = 
	- dH^{\rm{imag}}_{a,a} C_{a, b} 
	- C_{a, b} dH^{\rm{imag}}_{b, b} 
	+ 2 \sum_m (C_{a, m} dH^{\rm{imag}}_{m, m} C_{m, b}) ,
	\end{equation}
	then
	\begin{equation}
	\begin{aligned}
	\frac{dC_{b,a} dC_{a,b}}{Bdt} = 
	& \ \delta_{a,b} C_{b,a} C_{a,b} + C_{b,a} C_{a,b} - 2 C_{a,b} C_{b,b} C_{b,a} 
	+ C_{b,a} C_{a,b} + \delta_{a,b} C_{b,a} C_{a,b} -2 C_{a,a} C_{a,b} C_{b,a} \\
	& - 2 C_{b,a} C_{a,a} C_{a,b} - 2 C_{b,b} C_{b,a} C_{a,b}
	+ 4 \sum_m (C_{a,m} C_{m,b} C_{b,m} C_{m,a}) \\
	= & \
	2|C_{a,b}|^2 + 2\delta_{a,b} |C_{a,b}|^2 - 4 |C_{a,b}|^2 (C_{a,a}+C_{b,b})
	+ 4 \sum_m (|C_{a,m}|^2 |C_{b,m}|^2) .
	\end{aligned}
	\end{equation}
	
	For the second derivative, we have
	\begin{equation}
	\begin{aligned}
	- \sum_{m} (
	dH^{\rm{imag}}_{a,m} dC_{m,b}
	) = & \ - \sum_m [
	- dH^{\rm{imag}}_{a,m} dH^{\rm{imag}}_{m,m} C_{m,b}
	- dH^{\rm{imag}}_{a,m} C_{m,b} dH^{\rm{imag}}_{b,b}
	+ 2 \sum_n (C_{m,n} C_{n,b} dH^{\rm{imag}}_{a,m} dH^{\rm{imag}}_{n,n})
	] \\
	= & \
	Bdt(C_{a,b} + \delta_{a,b} C_{b,b} - 2 C_{a,a} C_{a,b}) , \\
	\\
	- \sum_{m} (
	dC_{a,m} dH^{\rm{imag}}_{m,b} 
	) = & \ - \sum_m [
	- dH^{\rm{imag}}_{a,a} C_{a,m} dH^{\rm{imag}}_{m,b} 
	- C_{a,m} dH^{\rm{imag}}_{m,m} dH^{\rm{imag}}_{m,b}
	+ 2 \sum_n (C_{a,n} dH^{\rm{imag}}_{n,n} C_{n,m} dH^{\rm{imag}}_{m,b})
	] \\
	= & \
	Bdt(\delta_{a,b} C_{b,b} + C_{a,b} - 2 C_{a,b} C_{b,b}) , \\
	\\
	2\sum_{m,n} (
	dC_{a,m} dH^{\rm{imag}}_{m,n} C_{n,b}
	) = & \ 2 \sum_{m,n} [
	- d H_{a,a} C_{a,m} dH_{m,n} C_{n,b}
	- C_{a,m} dH_{m,m} dH_{m,n} C_{n,b}
	+ 2 \sum_k (C_{a,k} dH_{k,k} C_{k,m} dH_{m,n} C_{n,b})
	] \\
	= & \
	Bdt[- 2 C_{a,a} C_{a,b} - 2 \sum_m (C_{a,m} C_{m,b}) + 4 \sum_m (C_{a,m} C_{m,m} C_{m,b})] , \\
	\\
	2\sum_{m,n} (
	C_{a,m} dH^{\rm{imag}}_{m,n} dC_{n,b}
	) = & \ 2 \sum_{m,n} [
	- C_{a,m} dH^{\rm{imag}}_{m,n} dH^{\rm{imag}}_{n,n} C_{n,b}
	- C_{a,m} dH^{\rm{imag}}_{m,n} C_{n,b} dH_{b,b}
	+ 2 \sum_k (C_{a,m} dH_{m,n} C_{n,k} dH_{k,k} C_{k,b})
	] \\
	= & \
	Bdt[-2\sum_m(C_{a,m}C_{m,b}) -2 C_{a,b} C_{b,b} + 4 \sum_m (C_{a,m} C_{m,m} C_{m,b})] ,
	\end{aligned}
	\end{equation}
	which leads to 
	\begin{equation}
	\frac{d^2 C_{a,b}}{Bdt} = 2 C_{a,b} + 2C_{b,b}\delta_{a,b} -4C_{a,b}(C_{a,a}+C_{b,b}) - 4 \sum_m (C_{a,m} C_{m,b}) + 8 \sum_m (C_{a,m} C_{m,m} C_{m,b}) .
	\end{equation}
	
	Thus, we have
	\begin{equation}
	\begin{aligned}
	& \frac{d^2C_{b,a} C_{a,b} + C_{b,a} d^2C_{a,b} }{2Bdt} \\
	= \ & \frac{1}{2} [
	2C_{b,a} C_{a,b} + 2 C_{b,b} C_{b,b} \delta_{a,b}
	- 4 C_{b,a} C_{a,b} (C_{a,a} + C_{b,b})
	- 4 \sum_m (C_{b,m} C_{m,a} C_{a,b})
	+ 8 \sum_m (C_{b,m}C_{m,m}C_{m,a}C_{a,b})
	\\
	& \quad
	+ 2C_{a,b} C_{b,a} + 2 C_{b,b} C_{b,b} \delta_{a,b}
	- 4 C_{b,a} C_{a,b} (C_{a,a} + C_{b,b})
	- 4 \sum_m (C_{b,a} C_{a,m} C_{m,b})
	+ 8 \sum_m (C_{b,a}C_{a,m}C_{m,m}C_{m,b})
	] \\
	= \ &
	2|C_{a,b}|^2 + 2|C_{b,b}|^2 \delta_{a,b}
	-4 |C_{a,b}|^2(C_{a,a}+C_{b,b})
	-2 \sum_m (C_{b,m}C_{m,a}C_{a,b} + C_{b,a}C_{a,m}C_{m,b}) \\
	& 
	+4 \sum_m (C_{b,m}C_{m,m}C_{m,a}C_{a,b} + C_{b,a}C_{a,m}C_{m,m}C_{m,b}) .
	\end{aligned}
	\end{equation}
	
	Finally, we get the differential equation for the imaginary dynamics as
	\begin{equation}
	\begin{aligned}
	\frac{d|C_{a,b}|^2}{Bdt} = & \
	2|C_{a,b}|^2 + 2\delta_{a,b} |C_{a,b}|^2 - 4 |C_{a,b}|^2 (C_{a,a}+C_{b,b})
	+ 4 \sum_m (|C_{a,m}|^2 |C_{b,m}|^2) \\
	& \
	+ 2|C_{a,b}|^2 + 2|C_{b,b}|^2 \delta_{a,b}
	-4 |C_{a,b}|^2(C_{a,a}+C_{b,b})
	-2 \sum_m (C_{b,m}C_{m,a}C_{a,b} + C_{b,a}C_{a,m}C_{m,b}) \\
	& \ 
	+4 \sum_m (C_{b,m}C_{m,m}C_{m,a}C_{a,b} + C_{b,a}C_{a,m}C_{m,m}C_{m,b}) \\
	= & \
	4|C_{a,b}|^2 + 4|C_{b,b}|^2 \delta_{a,b}
	-8 |C_{a,b}|^2(C_{a,a} + C_{b,b})
	+4 \sum_m (|C_{a,m}|^2 |C_{b,m}|^2) \\
	& \
	- 2 \sum_m (C_{b,m}C_{m,a}C_{a,b} + C_{b,a}C_{a,m}C_{m,b})
	+4 \sum_m (C_{b,m}C_{m,m}C_{m,a}C_{a,b} + C_{b,a}C_{a,m}C_{m,m}C_{m,b}) . \\
	\end{aligned}
	\end{equation}
	
	We can calculate the distribution function of the squared two-point correlation function as the average
	\begin{equation}
	f_{w,z} \equiv
	\begin{cases}
	\frac{1}{N}\sum_a (|C_{a,a}|^2) ,
	&  {\rm{when}} \ w = 0, z = 0 \\
	\frac{1}{N}\sum_a (|C_{a,a+w\delta x}|^2 + |C_{a,a-w\delta x}|^2) ,
	&  {\rm{when}} \ w > 0, z = 0 \\
	\frac{1}{N}\sum_a (|C_{a,a+z\delta y}|^2 + |C_{a,a-z\delta y}|^2) ,
	&  {\rm{when}} \ w = 0, z > 0 \\
	\frac{1}{N} \sum_{a} (
	|C_{a,a+w\delta x+z\delta y}|^2
	+ |C_{a,a+w\delta x-z\delta y}|^2 
	+ |C_{a,a-w\delta x+z\delta y}|^2
	+ |C_{a,a-w\delta x-z\delta y}|^2
	) ,
	& {\rm{when}} \ w > 0, z > 0 .
	\end{cases}
	\end{equation}
	After some straightforward algebra, for unitary dynamics, we have
	\begin{equation}
	\begin{aligned}
	\frac{1}{A} \partial_t f_{1,0} = 2f_{2,0} + 2f_{1,1} - 8f_{1,0} + \mu_x,
	\qquad \qquad w=1, z=0 \\
	\frac{1}{A} \partial_t f_{0,1} = 2f_{1,1} + 2f_{0,2} - 8f_{0,1} + \mu_y,
	\qquad \qquad w=0, z=1 \\
	\frac{1}{A} \partial_t f_{w,0} = 2f_{w+1,0} + 2f_{w-1,0} + 2f_{w,1} - 8f_{w,0},
	\qquad \qquad w>1, z=0 \\
	\frac{1}{A} \partial_t f_{0,z} = 2f_{1,z} + 2f_{0,z+1} + 2f_{0,z-1} - 8f_{0,z},
	\qquad \qquad w=0, z>1 \\
	\frac{1}{A} \partial_t f_{w,z} = 
	2f_{w+1,z} + 2f_{w-1,z} + 2f_{w,z+1} + 2f_{w,z-1} - 8f_{w,z} , 
	\qquad \qquad w \ge 1, z \ge 1 \\
	\end{aligned}
	\end{equation}
	with the source terms
	\begin{equation}
	\mu_x = \frac{1}{N}\sum_a [(C_{a,a} - C_{a+\delta x})^2 + (C_{a,a} - C_{a-\delta x})^2], \qquad
	\mu_y = \frac{1}{N}\sum_a [(C_{a,a} - C_{a+\delta y})^2 + (C_{a,a} - C_{a-\delta y})^2] .
	\end{equation}
	Here we note that the source term is constrained by: 1. the isotropy of the master equation; 2. the absence of the zero point $(0,0)$ in the lattice. This means that: 1. $\mu = \mu_x = \mu_y$; 2. if we take spatial continuum of the master equation, there should be only one source term $\mu$ that located at $(0,0)$. \\
	For imaginary dynamics, when $w > 1, z > 1$ we have
	\begin{equation}
	\begin{aligned}
	\frac{1}{B} \partial_t f_{w,z} = 
	& - 8f_{w,z} \sum_{m=1}^{\infty} \sum_{n=1}^{\infty} f_{m,n}
	- 8f_{w,z} \sum_{m=1}^{\infty} f_{m, 0} - 8f_{w,z} \sum_{n=1}^{\infty} f_{0, n} \\
	& + 4\sum_{m=1}^{\infty} \sum_{n=1}^{\infty} f_{m,n} f_{w+m,z+n}
	+ 4\sum_{m=1}^{\infty} f_{m,0} f_{w+m, z}
	+ 4\sum_{n=1}^{\infty} f_{0,n} f_{w, z+n} \\
	& + 2 \sum_{m=1}^{w-1} \sum_{n=1}^{z-1} f_{m,n} f_{w-m,z-n}
	+ 2 \sum_{m=1}^{w-1} (f_{m,0} f_{w-m,z} + f_{m,z} f_{w-m,0})
	+ 2 \sum_{n=1}^{z-1} (f_{0,n} f_{w,z-n} + f_{w,n} f_{0,z-n}) ;
	\\
	\end{aligned}
	\end{equation}
	when $w = 0, z = 1$ or $w = 1, z = 0$ or $w=1,z=1$, we have
	\begin{equation}
	\begin{aligned}
	\frac{1}{B} \partial_t f_{w,z} = 
	& - 8f_{w,z} \sum_{m=1}^{\infty} \sum_{n=1}^{\infty} f_{m,n}
	- 8f_{w,z} \sum_{m=1}^{\infty} f_{m, 0} - 8f_{w,z} \sum_{n=1}^{\infty} f_{0, n} \\
	& + 4\sum_{m=1}^{\infty} \sum_{n=1}^{\infty} f_{m,n} f_{w+m,z+n}
	+ 4\sum_{m=1}^{\infty} f_{m,0} f_{w+m, z}
	+ 4\sum_{n=1}^{\infty} f_{0,n} f_{w, z+n} ;
	\\
	\end{aligned}
	\end{equation}
	when $w = 0, z > 1$, we have
	\begin{equation}
	\begin{aligned}
	\frac{1}{B} \partial_t f_{w,z} = 
	& - 8f_{w,z} \sum_{m=1}^{\infty} \sum_{n=1}^{\infty} f_{m,n}
	- 8f_{w,z} \sum_{m=1}^{\infty} f_{m, 0} - 8f_{w,z} \sum_{n=1}^{\infty} f_{0, n} \\
	& + 4\sum_{m=1}^{\infty} \sum_{n=1}^{\infty} f_{m,n} f_{w+m,z+n}
	+ 4\sum_{m=1}^{\infty} f_{m,0} f_{w+m, z}
	+ 4\sum_{n=1}^{\infty} f_{0,n} f_{w, z+n} \\
	& + 2 \sum_{n=1}^{z-1} f_{0,n} f_{w,z-n} ;
	\\
	\end{aligned}
	\end{equation}
	when $w > 1, z = 0$, we have
	\begin{equation}
	\begin{aligned}
	\frac{1}{B} \partial_t f_{w,z} = 
	& - 8f_{w,z} \sum_{m=1}^{\infty} \sum_{n=1}^{\infty} f_{m,n}
	- 8f_{w,z} \sum_{m=1}^{\infty} f_{m, 0} - 8f_{w,z} \sum_{n=1}^{\infty} f_{0, n} \\
	& + 4\sum_{m=1}^{\infty} \sum_{n=1}^{\infty} f_{m,n} f_{w+m,z+n}
	+ 4\sum_{m=1}^{\infty} f_{m,0} f_{w+m, z}
	+ 4\sum_{n=1}^{\infty} f_{0,n} f_{w, z+n} \\
	& + 2 \sum_{m=1}^{w-1} f_{m,0} f_{w-m,z} ;
	\\
	\end{aligned}
	\end{equation}
	when $w = 1, z > 1$, we have
	\begin{equation}
	\begin{aligned}
	\frac{1}{B} \partial_t f_{w,z} = 
	& - 8f_{w,z} \sum_{m=1}^{\infty} \sum_{n=1}^{\infty} f_{m,n}
	- 8f_{w,z} \sum_{m=1}^{\infty} f_{m, 0} - 8f_{w,z} \sum_{n=1}^{\infty} f_{0, n} \\
	& + 4\sum_{m=1}^{\infty} \sum_{n=1}^{\infty} f_{m,n} f_{w+m,z+n}
	+ 4\sum_{m=1}^{\infty} f_{m,0} f_{w+m, z}
	+ 4\sum_{n=1}^{\infty} f_{0,n} f_{w, z+n} \\
	& + 2 \sum_{n=1}^{z-1} (f_{0,n} f_{w,z-n} + f_{w,n} f_{0,z-n}) ;
	\\
	\end{aligned}
	\end{equation}
	when $w > 1, z = 1$, we have
	\begin{equation}
	\begin{aligned}
	\frac{1}{B} \partial_t f_{w,z} = 
	& - 8f_{w,z} \sum_{m=1}^{\infty} \sum_{n=1}^{\infty} f_{m,n}
	- 8f_{w,z} \sum_{m=1}^{\infty} f_{m, 0} - 8f_{w,z} \sum_{n=1}^{\infty} f_{0, n} \\
	& + 4\sum_{m=1}^{\infty} \sum_{n=1}^{\infty} f_{m,n} f_{w+m,z+n}
	+ 4\sum_{m=1}^{\infty} f_{m,0} f_{w+m, z}
	+ 4\sum_{n=1}^{\infty} f_{0,n} f_{w, z+n} \\
	& + 2 \sum_{m=1}^{w-1} (f_{m,0} f_{w-m,z} + f_{m,z} f_{w-m,0}) .
	\\
	\end{aligned}
	\end{equation}

The equation of motion for $f_{w,z}$ during the mixed nonunitary time evolution is just a combination of the derived real and imaginary dynamics.
For general case (when $w > 1, z > 1$), it is
\begin{equation}\label{eq:2dMasterEq}
\begin{aligned}
\partial_t f_{w,z} = \
& \theta (f_{w+1,z} + f_{w-1,z} + f_{w,z+1} + f_{w,z-1} - 4f_{w,z}) 
- 8f_{w,z} \sum_{m=1}^{\infty} \sum_{n=1}^{\infty} f_{m,n}
- 8f_{w,z} \sum_{m=1}^{\infty} f_{m, 0} - 8f_{w,z} \sum_{n=1}^{\infty} f_{0, n} \\
& + 4\sum_{m=1}^{\infty} \sum_{n=1}^{\infty} f_{m,n} f_{w+m,z+n}
+ 4\sum_{m=1}^{\infty} f_{m,0} f_{w+m, z}
+ 4\sum_{n=1}^{\infty} f_{0,n} f_{w, z+n} 
+ 2 \sum_{m=1}^{w-1} \sum_{n=1}^{z-1} f_{m,n} f_{w-m,z-n} \\
& + 2 \sum_{m=1}^{w-1} (f_{m,0} f_{w-m,z} + f_{m,z} f_{w-m,0})
+ 2 \sum_{n=1}^{z-1} (f_{0,n} f_{w,z-n} + f_{w,n} f_{0,z-n}) ,
\\
\end{aligned}
\end{equation}
where $\theta \sim \frac{A}{B}$ is a constant to describe the strength of the real time evolution in the mixed nonunitary dynamics.
This equation describes the spreading of the squared correlation function in the bulk of the system.
Here the terms with prefactor $\theta$ come from the unitary dynamics, and the rest terms from the imaginary dynamics.
In Figs.~\ref{fig:simulation_pde_early} and~\ref{fig:simulation_pde}, we plot the numerical results of the large time behavior of the distribution function $f_{1,r}$ for two-points on the square lattice with distance $r=\sqrt{w^2+z^2}$.
For different settings of the strength of the real time evolution $\theta$, $f_{r}$ exhibits universal scaling behavior close to that of Eq.~\eqref{eq:CorrScaling}.

\end{widetext}

\begin{figure}\centering
	\includegraphics[width=\columnwidth]{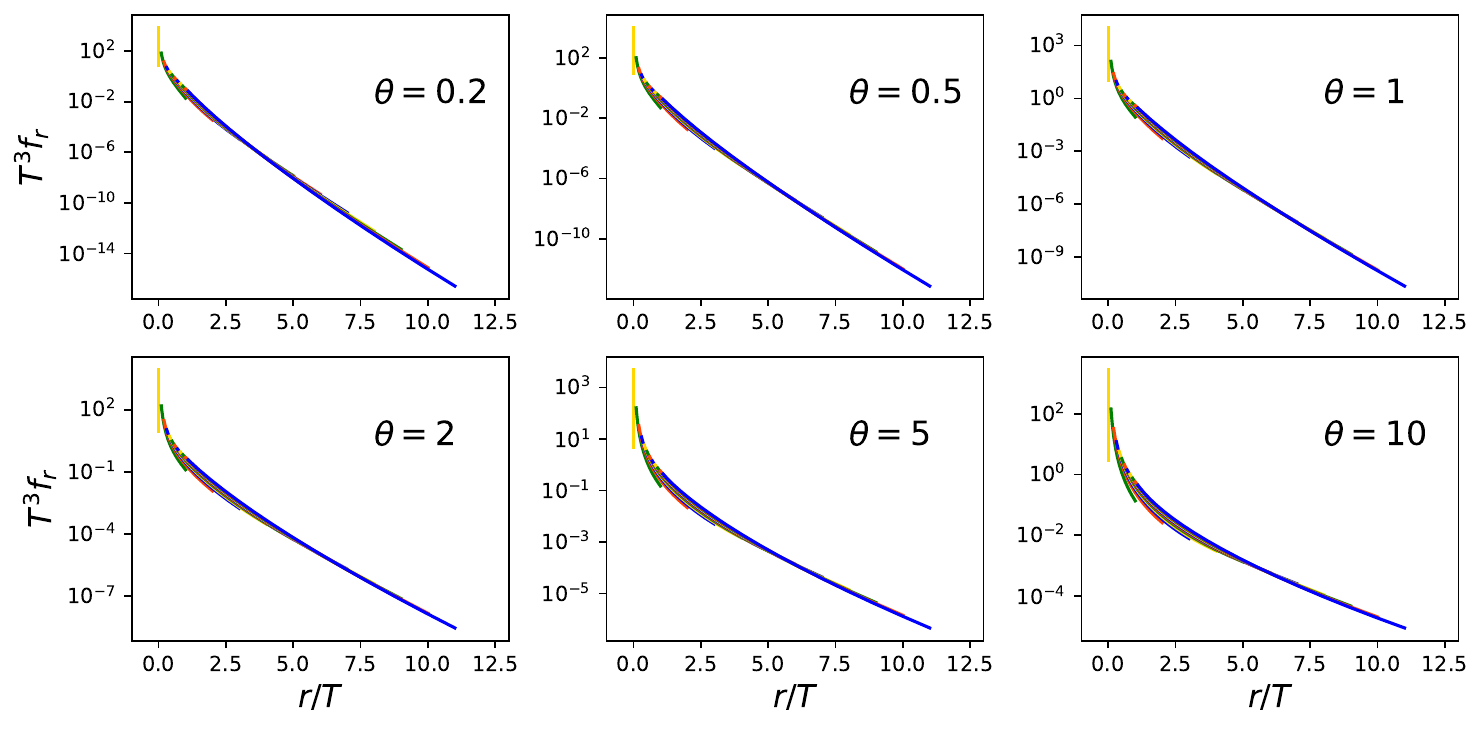}
	\caption{
		\label{fig:simulation_pde_early}
		Numerical simulation of the equation of motion for $f_{w,z}$ during the mixed nonunitary time evolution, for different values of $\theta \sim \frac{A}{B}$.
		The presented data are for time $t \in [5, 50]$, and the total system size is set to be $100 \times 100$.
	}
\end{figure}

\begin{figure}\centering
	\includegraphics[width=\columnwidth]{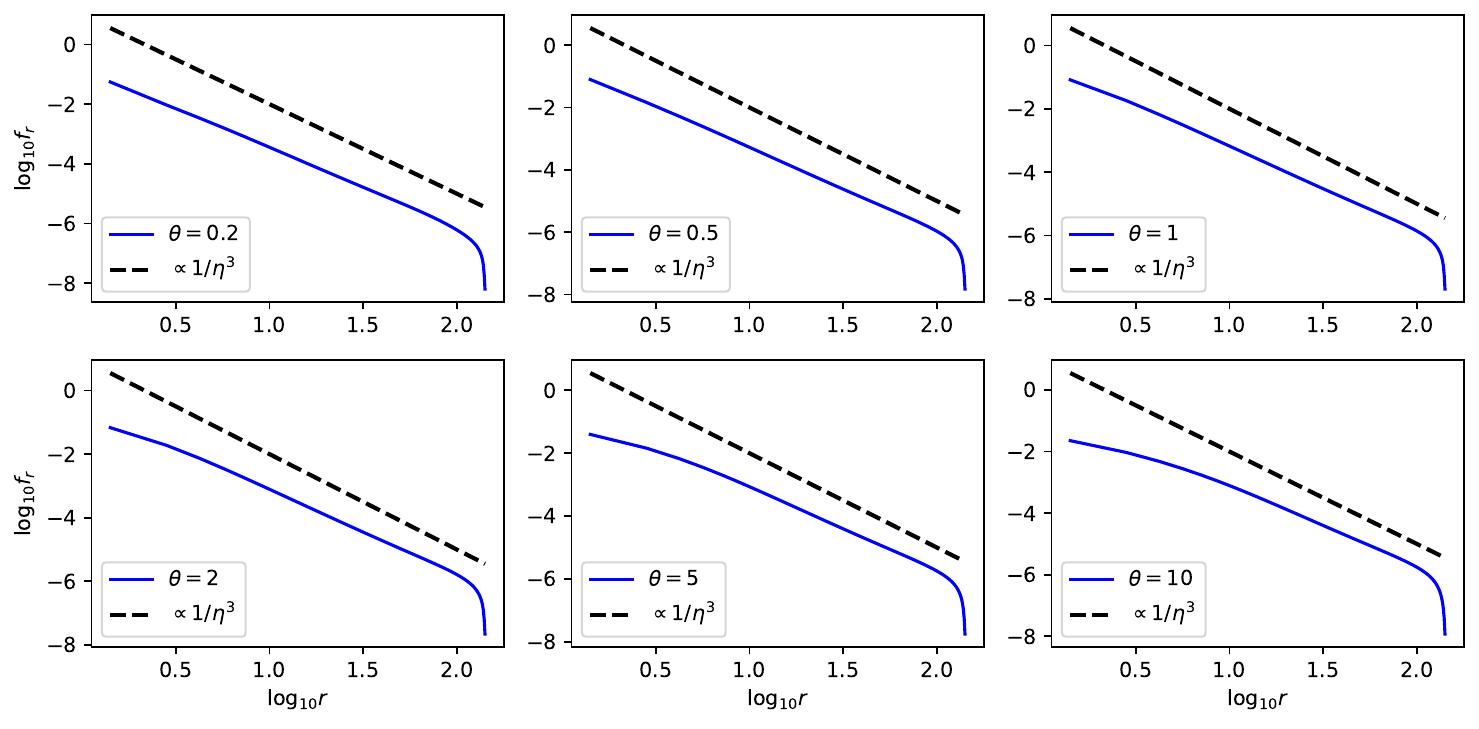}
	\caption{
		\label{fig:simulation_pde}
		Numerical simulation of the equation of motion for $f_{w,z}$ during the mixed nonunitary time evolution, for different values of $\theta \sim \frac{A}{B}$.
		The presented data are for time $t = 200$, which already reaches the steadiness for the chosen total system size $100 \times 100$.
	}
\end{figure}


\begin{widetext}

To make analytical argument on the steady solution of the above discrete differential equation, we take the right part of the master in the continuum limit
and let $f_{(-w, z)} = f_{(w, -z)} = f_{(-w, -z)} = f_{(w, z)}$.
For the terms from unitary dynamics, we have
\begin{equation}
f_{w+1,z} + f_{w-1,z} + f_{w,z+1} + f_{w,z-1} - 4f_{w,z}
= \frac{f_{w+1,z} - 2f_{w,z} + f_{w-1,z}}{1} + \frac{f_{w,z+1} + f_{w,z-1} - 2f_{w,z}}{1} .
\end{equation}
The two terms in the right hand side of the above equation have the form of the second derivative with spatial spacing $\Delta x = 1$ (the lattice constant). In the continuum limit, it becomes the diffusion term
\begin{equation}
f_{w+1,z} + f_{w-1,z} + f_{w,z+1} + f_{w,z-1} - 4f_{w,z} 
\qquad \Longrightarrow \qquad \nabla^2 f_{w,z} .
\end{equation}

The terms from imaginary dynamics can be separated into two parts. The first one is the terms proportional to $f_{w,z}$
with the prefactor as the summation of distribution function $f_{w,z}$ in the full space.
It is important to note that this summation is required to be positive (it is the value of squared two-point correlation) and in the order of $\mathcal{O}(1)$ (the integral of probability distribution should be able to normalized to 1).
Therefore, we can write it into
\begin{equation}
- 8 f_{w,z} \left(
	\sum_{m=1}^{\infty} \sum_{n=1}^{\infty} f_{m,n} 
	+ \sum_{m=1}^{\infty} f_{m, 0}
	+ \sum_{n=1}^{\infty} f_{0, n}
	\right)  
\qquad \Longrightarrow \qquad - c f_{w,z}
\end{equation}
with a positive $\mathcal{O}(1)$ constant $c$.

The rest nonlinear terms can be combined in the form of convolution as
\begin{equation}
\begin{aligned}
& 4\sum_{m=1}^{\infty} \sum_{n=1}^{\infty} f_{m,n} f_{w+m,z+n}
+ 4\sum_{m=1}^{\infty} f_{m,0} f_{w+m, z}
+ 4\sum_{n=1}^{\infty} f_{0,n} f_{w, z+n} \\
& + 2 \sum_{m=1}^{w-1} \sum_{n=1}^{z-1} f_{m,n} f_{w-m,z-n} 
+ 2 \sum_{m=1}^{w-1} (f_{m,0} f_{w-m,z} + f_{m,z} f_{w-m,0})
+ 2 \sum_{n=1}^{z-1} (f_{0,n} f_{w,z-n} + f_{w,n} f_{0,z-n}) \\
= & \ 2\sum_{m=0}^{\infty} \sum_{n=0}^{\infty} f_{m,n} f_{w+m,z+n} 
+ 2\sum_{m=w}^{\infty} \sum_{n=z}^{\infty} f_{m,n} f_{m-w,n-z} 
+ 2 \sum_{m=0}^{w-1} \sum_{n=0}^{z-1} f_{m,n} f_{w-m,z-n} \\
= & \ 2\sum_{m=0}^{\infty} \sum_{n=0}^{\infty} f_{m,n} f_{w+m,z+n} 
+ 2\sum_{m=w}^{\infty} \sum_{n=z}^{\infty} f_{m,n} f_{w-m,z-n} 
+ 2 \sum_{m=0}^{w-1} \sum_{n=0}^{z-1} f_{m,n} f_{w-m,z-n} \\
= & \ 2\int_0^\infty dxdy f_{x, y} f_{w+x, z+y}
+ 2\int_0^\infty dxdy f_{x, y} f_{w-x, z-y} 
= 2\int_{-\infty}^0 dxdy f_{x, y} f_{w+x, z+y}
+ 2\int_0^\infty dxdy f_{x, y} f_{w-x, z-y} \\
= & \ 2\int_{-\infty}^\infty dxdy f_{x, y} f_{w+x, z+y} ,
\end{aligned}
\end{equation}
where we have assumed $f_{m,n} = f_{-m,n} = f_{m,-n} = f_{-m,-n}$, and the following relation is used in the first line:
\begin{equation}
\sum_{m=0}^{\infty} \sum_{n=0}^{\infty} f_{m,n} f_{w+m,z+n}
= \sum_{m=w}^{\infty} \sum_{n=z}^{\infty} f_{m,n} f_{m-w,n-z} .
\end{equation}

Based on the above mappings, finally we obtain the continuum version of the nonlinear master equation as
\begin{equation}
	\partial_t f_{w,z} = \theta \nabla^2 f_{w,z} + \mu\delta_{(w,z), (0,0)} - cf_{w,z} + \int_{-\infty}^{\infty} f_{x,y} f_{w-x, z-y} dxdy ,
\end{equation}
where $\mu$ and $c$ are both positive $\mathcal{O}(1)$ constants.
We have numerically confirmed that the steady solution with different values of $\theta$ gives universal scaling behavior.
This implies that the diffusive term can be ignored, i.e., we can set $\theta = 0$.
This leads to the equation of steady $f(r)$,
\begin{equation}
	0 = - cf_{w,z} + \int_{-\infty}^{\infty} f_{x,y} f_{w-x, z-y} dxdy ,
\end{equation}
where we have used the steadiness condition $\partial_t=0$.
In this case, the ansatz for solving the equation is
\begin{equation}
	f(r) \sim 1 / |r|^3.
\end{equation}
This solution can be tested by simply substituting it into the nonlinear master equation, then we have
\begin{equation}
	0 = - \frac{c}{|r'|^3} + \int_0^{2\pi} d\varphi \int_0^\infty r dr \frac{1}{|r|^3} \frac{1}{|r'-r|^3} .
\end{equation}
The last term of integral can be calculated as
\begin{equation}
	\begin{aligned}
		& \int_0^{2\pi} d\varphi \int_0^\infty r dr \frac{1}{|r|^3} \frac{1}{|r'-r|^3} \
		= \ \int_0^\infty dr \frac{1}{r^2} \int_0^{2\pi} d\varphi \frac{1}{[(r')^2 + r^2 - 2rr'\cos \varphi]^{3/2}} \\
		& {\rm{take}} \qquad t = \tan \frac{\varphi}{2}, \cos \varphi = \frac{1 - t^2}{1 + t^2} \\
		= \  &  \int_0^\infty dr \frac{1}{r^2} \int_0^{\infty} \frac{4dt}{[(r'-r)^2 + t^2 (r'+r)^2]^{3/2}} 
		= 4 \int_0^\infty dr \frac{1}{r^2} \frac{1}{(r'-r)^3}\int_0^{\infty} \frac{dt}{[1 + t^2 (\frac{r'+r}{r'-r})^2]^{3/2}} \\
		{\rm{take}} & \qquad \tan s =  t (\frac{r'+r}{r'-r}) \\
		= \  &  4 \int_0^\infty \frac{dr}{r^2 (r'-r)^3}\int_0^{\infty} \frac{d \tan s / \frac{r'+r}{r'-r}}{(1 + \tan^2 s)^{3/2}}
		= 4 \int_0^\infty \frac{dr}{r^2 (r'-r)^2 (r'+r)}\int_0^{\infty} \frac{d \tan s}{(1 + \tan^2 s)^{3/2}} \\
		= \  &  4 \int_0^\infty \frac{dr}{r^2 (r'-r)^2 (r'+r)} \int_0^{\pi/2} \frac{ds / \cos^2 s}{(1 + \frac{\sin^2 x}{\cos^2 x})^{3/2}} 
		= 4 \int_0^\infty \frac{dr}{r^2 (r'-r)^2 (r'+r)} \int_0^{\pi/2} ds \frac{1}{\cos^2 s} \frac{1}{(1/\cos^2 x)^{3/2}} \\
		= \  &  4 \int_0^\infty \frac{dr}{r^2 (r'-r)^2 (r'+r)}  \int_0^{\pi/2} ds \cos s
		= 4 \int_0^\infty \frac{dr}{r^2 (r'-r)^2 (r'+r)} . \\
	\end{aligned}
\end{equation}
The above integral is divergent, and need to introduce UV cutoff for  solving it, as 
\begin{equation}
	\int_\epsilon^\infty \frac{dr}{r^2 (r'-r)^2 (r'+r)} \sim \int_1^\infty \frac{dr}{r^2 (r'-r)^2 (r'+r)} \sim \frac{1}{|r'|^3} .
\end{equation}
This is consistent with our ansatz.

\section{Quantum entanglement from the quasiparticle picture}\label{app:quasi}
In this appendix we introduce the quasiparticle picture that was mentioned in the main text.
It was proposed by Skinner and Nahum~\cite{nahum2020entanglement} to describe a Majorana dynamics with diffusion-annihilation process.
In this paper, we show that this picture provides a way to estimate the entanglement entropy in the mixed nonunitary dynamics of free fermions.
The quantum entanglement in this picture is considered to be produced by quasiparticle pairs in the system, and the scaling behavior of the entanglement entropy is determined by the distribution function of those pairs.

We first review the case of the $(1+1)$D nonunitary dynamics, where the mutual information between two points  has $I({\rm points}) \propto 1/r^2$ scaling and the entanglement entropy exhibits the typical logarithmic growth as $S(L_A) \propto \ln L_A$ for $(1+1)$D critical systems.
Below we show that these scaling behaviors are consistent in the quasiparticle picture.
Let us define the probability of two specific particles with distance $r$ paired to be $P(r)$, then the mutual information between two subsystems $A$ and $B$ is 
\begin{equation}{\label{eq:summation_EE}}
I_{A,B} = \sum_{\{ \rm{point \ in \ A} \}} \sum_{\{ \rm{point \ in \ B} \}} P(\rm{distance \ between \ the \ point \ in \ A \ and \ the \ point \ in \ B}),
\end{equation}
when $A$ and $B$ are complementary to the total system, this becomes entanglement entropy.

If $A$ and $B$ are just two points in the system, we have the mutual information between those two points as
\begin{equation}
I_{A,B} = P(r_{A,B}) \propto \frac{1}{r_{A,B}^a} ,
\end{equation}
where we have assumed the power-law decay of mutual information.
Since $I({\rm points}) \propto 1/r^2$, we have $a = 2$ in $(1+1)$D nonunitary dynamics.

For the case that an infinite total system is bipartite into two subsystem $A$ and $B$, 
Eq.~\ref{eq:summation_EE} becomes
\begin{equation}
S_A = \sum_{x_A = 1}^{L_A} \ \sum_{r=L_A-x_A+1}^{\infty} P(r).
\end{equation}
The summation can be solved analytically when $a = 2$, it gives 
\begin{equation}
S_A \sim \ln L_A ,
\end{equation}
i.e., the observed critical entanglement scaling is correctly obtained from the scaling of mutual information.

Moreover, we find that these observations can be extended into higher dimensions, namely, the $(2+1)$D mixed nonunitary dynamics studied in this work.
From large-scale numerical simulations, we obtain $a = 3$ for the $(2+1)$D case.
The problem is that the summation in $(2+1)$D cannot be calculated analytically.
To calculate it, we need to do continuum extension for the summation, as 
\begin{equation}
S_{A} \sim \int_A dV_A \int_B dV_B P(r_{A,B}) = \int_A dV_A \int_B dV_B \frac{1}{r_{A,B}^a} ,
\end{equation}
where $A$ and $B$ are complementary to the total system.

For simplicity, we consider the subsystem with disc geometry, then the above integral becomes
\begin{equation}
S_{A} \sim \int_0^{L_A} r' dr' \int_0^{2\pi} d\varphi' \int_{L_A+c}^{\infty} r dr \int_0^{2\pi} d\varphi \frac{1}{|\vec{r} - \vec{r}'|^a} ,
\end{equation}
where $c$ is the lattice constant. It is obvious that $a$ should be larger than 2 (the dimension) to make the distribution function normalizable.

It should be mentioned that a direct solution of the integral with $a=3$ will lead to infinity, i.e. we must impose a UV cutoff to get a convergent result of the asymptotic behavior at $L_A \to \infty$.
Let us first rescale the length variables as $r' = uL_A$ and $r = vL_A$, and then the integral becomes
\begin{equation}
\begin{aligned}
S_{A} \sim \
& L_A \int_0^{1} u du \int_0^{2\pi} d\varphi' \int_{1+\epsilon}^{\infty} v dv \int_0^{2\pi} d\varphi \frac{1}{|\vec{u} - \vec{v}|^3} \\
&  =  \  L_A \int_0^{1} u du \int_0^{2\pi} d\varphi' \int_{1+\epsilon}^{\infty} v dv \int_0^{2\pi} d\varphi \frac{1}{(u^2+v^2-2uv\cos \varphi)^{3/2}} \\
&  = \  2\pi L_A \int_{1+\epsilon}^{\infty} v dv W ,  
\end{aligned}
\end{equation}
with
\begin{equation}
\begin{aligned}
W = \
& \int_0^{1} u du \int_0^{2\pi} d\varphi  \frac{1}{(u^2+v^2-2uv\cos \varphi)^{3/2}} \\
{\rm{take}} & \qquad t = \tan \frac{\varphi}{2}, \cos \varphi = \frac{1 - t^2}{1 + t^2} \\
= \  &  \int_0^{1} u du \int_0^{\infty} \frac{4dt}{[(u-v)^2 + t^2 (u+v)^2]^{3/2}} 
= 4 \int_0^{1} u du \frac{1}{(u-v)^3}\int_0^{\infty} \frac{dt}{[1 + t^2 (\frac{u+v}{u-v})^2]^{3/2}} \\
{\rm{take}} & \qquad \tan s =  t (\frac{u+v}{u-v}) \\
= \  &  4 \int_0^{1}  \frac{u du}{(u-v)^3}\int_0^{\infty} \frac{d \tan s / \frac{u+v}{u-v}}{(1 + \tan^2 s)^{3/2}}
= 4 \int_0^{1}  \frac{u du}{(u-v)^2(u+v)} \int_0^{\infty} \frac{d \tan s}{(1 + \tan^2 s)^{3/2}} \\
= \  &  4 \int_0^{1}  \frac{u du}{(u-v)^2(u+v)} \int_0^{\pi/2} \frac{ds / \cos^2 s}{(1 + \frac{\sin^2 x}{\cos^2 x})^{3/2}}
= 4 \int_0^{1}  \frac{u du}{(u-v)^2(u+v)} \int_0^{\pi/2} ds \frac{1}{\cos^2 s} \frac{1}{(1/\cos^2 x)^{3/2}} \\
= \  &  4 \int_0^{1}  \frac{u du}{(u-v)^2(u+v)} \int_0^{\pi/2} ds \cos s
= 4 \int_0^{1}  \frac{u du}{(u-v)^2(u+v)} 
= 2 \int_0^1 du \frac{1}{u-v} (\frac{1}{u-v} + \frac{1}{u+v}) \\
= \  &  2 \int_0^{1} du [\frac{1}{(u-v)^2} + \frac{1}{2v} (\frac{1}{u-v} - \frac{1}{u+v})]
= 2 \int_0^{1} \frac{du}{(u-v)^2} + \frac{1}{v} \int_0^{1} \frac{du}{u-v} - \frac{1}{v} \int_0^{1} \frac{du}{u+v} \\
= \  &  2(\frac{-1}{1-v} - \frac{-1}{-v}) + \frac{1}{v}[\ln v - \ln (v-1)] - \frac{1}{v}[\ln (1+v) - \ln (v)]
=  \frac{1}{v} [\frac{2}{v-1} + 2\ln v - \ln(v+1) - \ln(v-1)] .
\end{aligned}
\end{equation}
Then we have
\begin{equation}
S_{A} \sim \
2\pi L_A \int_{1+\epsilon}^{\infty} dv [\frac{2}{v-1} + 2\ln v - \ln(v+1) - \ln(v-1)] ,
\end{equation}
with $\varepsilon = {c}/{L_A}$.
After the rescale of the unit of $L_A$, the limit $L_A \to \infty$ becomes $\varepsilon \to 0$.

To get the convergent asymptotic behavior, we take the scale of the total system size to be a finite value $b = L/L_A$, i.e.
\begin{equation}
\begin{aligned}
S_{A} \sim \
& 2\pi L_A \int_{1+\epsilon}^{b} dv [\frac{2}{v-1} +  2\ln v - \ln(v+1) - \ln(v-1)] \\
& = \ 
4\pi L_A [\ln (b-1) - \ln\varepsilon]
+ 4\pi L_A [b\ln b - b - (1+\varepsilon) \ln(1+\varepsilon) + (1+\varepsilon)] \\
& \quad - 2\pi L_A [ (b+1)\ln(b+1) - (b+1) - (2+\varepsilon)\ln(2+\varepsilon) + (2+\varepsilon) ] \\
& \quad - 2\pi L_A [ (b-1)\ln(b-1) - (b-1) - \varepsilon\ln\varepsilon + \varepsilon ] . \\
\end{aligned}
\end{equation}
Take the limit $\varepsilon \to 0$, we have
\begin{equation}
\begin{aligned}
S_{A} \sim \
& 4\pi L_A [\ln (b-1) - \ln(\varepsilon)]
+ 4\pi L_A [b\ln b - b - 1] \\
& \ - 2\pi L_A [ (b+1)\ln(b+1) - (b+1) - 2\ln2 + 2 ] 
- 2\pi L_A [ (b-1)\ln(b-1) - (b-1) ] , \\
\end{aligned}
\end{equation}
ignoring the terms that do not depend on $L_A$, we get
\begin{equation}
S_{A} \sim \ - 4\pi L_A \ln\varepsilon = - 4\pi L_A \ln \frac{a}{L_A} = 4\pi L_A \ln L_A  - 4\pi L_A \ln a \sim 4\pi L_A \ln L_A ,
\end{equation}
such that we connect the $\propto 1/r^3$ correlation with the $L_A\ln L_A$ entanglement entropy in the quasiparticle picture, as the same as the observation in the numerical simulation of the mixed nonunitary dynamics.

One more evidence is the mutual information between two strips $A$ and $B$ with size $L_A \times L_y$ ($L_A \ll L_x$ and $L_A \ll L_y$)
\begin{equation}
\begin{aligned}
I({\rm{strip}}) \sim &
\int_0^{L_A} dx_A \int_r^{r+L_A} dx_B \int_0^{L_y} dy_A \int_0^{L_y} dy_B [(x_A - x_B)^2 + (y_A-y_B)]^{-3/2} \\
\sim & 2 L_A^2 L_y \int_{0}^{L/2} dl_y [r^2 + l_y^2]^{-3/2}
= \frac{2L_y}{r^2} \sin \arctan \frac{L_y}{2r} \propto \frac{L_y}{r^2}.
\end{aligned}
\end{equation}
Here we have used the periodic boundary condition.
The $1/r^2$ scaling of $I({\rm{strip}})$ is also obtained in the numerics.
Thus, we conclude that the entanglement scaling of the steady state of the mixed nonunitary dynamics can be described by the quasiparticle picture.

\section{Extension of the analytical results into general $(d+1)$-dimensional random nonunitary dynamics of free fermions}\label{app:higher}

In this appendix, we extend the analytical results into general $d+1$ dimension.
To achieve this, we make the following conjecture: in any spatial dimension, the scaling behavior of the steady-state correlations can be captured by the nonlinear master equation, and the entanglement structure is described by the quasiparticle picture.
After this, the problem turns into the following two parts: 1. solve the nonlinear master equation in general $d+1$ dimensions; 2. substitute the form of correlations into the quasiparticle picture to get the entanglement entropy.
We find that for $(d+1)$-dimensional dynamics, the steady state has $\propto 1/r^{d+1}$ scaling two-point correlations, and $\propto L^{d-1}\log L$ entanglement entropy.
Moreover, the solution of the nonlinear master equation implies that the chaotic unitary background dynamics is not necessary for reaching the steady state with special critical entanglement structure.

We begin with the nonlinear master equation. 
In $d+1$ dimensions, the equation of motion of the two-point correlation function during the unitary and imaginary Brownian dynamics should have the following form:
\begin{equation}\label{BrownianEq}
\begin{aligned}
\frac{d|C_{a,b}|^2}{Adt} \quad =  & \quad
\sum_{i=1}^d 
\Big\{
\left[ |C_{a-\delta x_i, b}|^2 + |C_{a+\delta x_i, b}|^2 + |C_{a, b-\delta x_i}|^2 + |C_{a, b+\delta x_i}|^2 \right] \\
& \qquad \qquad -2C_{b, b} \left[ C_{b+\delta x_i, b+ \delta x_i} \delta_{a-\delta x_i,b} + C_{b-\delta x_i, b- \delta x_i} \delta_{a+\delta x_i,b} \right] \\
& \qquad \qquad + 4 |C_{a,b}|^2 - 2 \delta_{a,b} C_{b,b} \left[ C_{b+\delta x_i, b+ \delta x_i} +  C_{b-\delta x_i, b- \delta x_i}  \right] \ \Big\} \\
\\
\frac{d|C_{a,b}|^2}{Bdt} \quad = & \quad 
4|C_{a,b}|^2 + 4|C_{b,b}|^2 \delta_{a,b}
-8 |C_{a,b}|^2(C_{a,a} + C_{b,b})
+4 \sum_m (|C_{a,m}|^2 |C_{b,m}|^2) \\
& \
- 2 \sum_m (C_{b,m}C_{m,a}C_{a,b} + C_{b,a}C_{a,m}C_{m,b})
+4 \sum_m (C_{b,m}C_{m,m}C_{m,a}C_{a,b} + C_{b,a}C_{a,m}C_{m,m}C_{m,b}) \\
\end{aligned}
\end{equation}
The above equation is quite complicated. To simplify the problem, here we note that the terms in the last line $C_{b,m}C_{m,a}C_{a,b}$ and $C_{b,m}C_{m,m}C_{m,a}C_{a,b}$ can be ignored, since we expect their average to be zero during the  random dynamics.

Recall that we only care about the distribution of the squared two-point correlation function instead, which can be defined by the average
\begin{equation}
f_X \equiv 
\frac{1}{N} \sum_{a} (
|C_{a,a \pm X_1\delta x_1 \pm \dots \pm X_d\delta x_d}|^2
) ,
{\rm{when}} \ X_1 > 0, \dots \ , X_d > 0
\end{equation}
where $X = (X_1, \dots \ , X_d)$ is the vector between the considered two points, and the summation runs over all possibility of the combination of the plus or minus sign in the second index of $C$.
For any $X_i = 0$, the terms in direction $x_i$ vanish to avoid double counting, and when $X_1 = \dots \ = X_d = 0$ we have $f_X \equiv \frac{1}{N} |C_{a,a}|^2$.

We can write the equation of motion for the distribution function $f_X$ from Eq.~\eqref{BrownianEq}; after some straightforward algebra, we have
\begin{equation}\label{eq:MasterEq}
\begin{aligned}
\partial_t f_X = 
& \ \theta \left[ \sum_{i=1}^d \left( f_{X-\delta x_i} + f_{X+\delta x_i} - 2f_X \right) \right] - 8 f_X \sum_{x_1 = 0}^\infty \dots \sum_{x_d = 0}^\infty f_{(x_1, \dots , x_d)} + 8 f_X f_0 \\
& \ + 4 \sum_{x_1 = 0}^\infty \dots \sum_{x_d = 0}^\infty f_{(x_1, \dots , x_d)} f_{X + (x_1, \dots , x_d)} - 4 f_X f_0 
+ 2 \sum_{x_1 = 0}^{X_1} \dots \sum_{x_d = 0}^{X_d} f_{(x_1, \dots , x_d)} f_{X - (x_1, \dots , x_d)} - 4 f_X f_0 \\ 
= & \ \theta \left[ \sum_{i=1}^d \left( f_{X-\delta x_i} + f_{X+\delta x_i} - 2f_X \right) \right] - 8 f_X \sum_{x_1 = 0}^\infty \dots \sum_{x_d = 0}^\infty f_{(x_1, \dots , x_d)} \\
& \ + 4 \sum_{x_1 = 0}^\infty \dots \sum_{x_d = 0}^\infty f_{(x_1, \dots , x_d)} f_{X + (x_1, \dots , x_d)} + 2 \sum_{x_1 = 0}^{X_1} \dots \sum_{x_d = 0}^{X_d} f_{(x_1, \dots , x_d)} f_{X - (x_1, \dots , x_d)} 
\end{aligned}
\end{equation}

First, we point out that the term with prefactor $\theta$ comes from the unitary dynamics.
After taking the continuum limit in space, it can be written in the form of the second derivative, and plays the role of diffusive term.
Second, the term that is proportional to $f_X$ has a prefactor as the summation $\sum_{x_1 = 0}^\infty \dots \sum_{x_d = 0}^\infty f_{(x_1, \dots , x_d)}$ in the full space.
This summation can be treated as a $\mathcal{O}(1)$ constant, since the integral of probability distribution should be normalizable.
Third, we notice that the last line in Eq.~\eqref{eq:MasterEq} has the form of convolution. 
To see this, we apply the similar treatment as the $(2+1)$D case discussed in Appendix~\ref{app:master}, as
\begin{equation}
\begin{aligned}
& 4 \sum_{x_1 = 0}^\infty \dots \sum_{x_d = 0}^\infty f_{(x_1, \dots , x_d)} f_{X + (x_1, \dots , x_d)} + 2 \sum_{x_1 = 0}^{X_1} \dots \sum_{x_d = 0}^{X_d} f_{(x_1, \dots , x_d)} f_{X - (x_1, \dots , x_d)} \\
= \ &
2\sum_{x_1 = 0}^\infty \dots \sum_{x_d = 0}^\infty f_{(x_1, \dots , x_d)} f_{X + (x_1, \dots , x_d)}
+ 2\sum_{x_1 = X_1}^\infty \dots \sum_{x_d = X_d}^\infty f_{(x_1, \dots , x_d)} f_{(x_1, \dots , x_d) - X}
+ 2\sum_{x_1 = 0}^{X_1} \dots \sum_{x_d = 0}^{X_d} f_{(x_1, \dots , x_d)} f_{X - (x_1, \dots , x_d)} \\
= \ &
2\sum_{x_1 = 0}^\infty \dots \sum_{x_d = 0}^\infty f_{(x_1, \dots , x_d)} f_{X + (x_1, \dots , x_d)}
+ 2\sum_{x_1 = X_1}^\infty \dots \sum_{x_d = X_d}^\infty f_{X - (x_1, \dots , x_d)} f_{(x_1, \dots , x_d) }
+ 2\sum_{x_1 = 0}^{X_1} \dots \sum_{x_d = 0}^{X_d} f_{(x_1, \dots , x_d)} f_{X - (x_1, \dots , x_d)} \\
= \ &
2\sum_{x_1 = 0}^\infty \dots \sum_{x_d = 0}^\infty f_{(x_1, \dots , x_d)} f_{X + (x_1, \dots , x_d)}
+ 2\int_0^{\infty} dx^d f_x f_{X-x}  \
= \ 
2\int_0^\infty dx^d f_x f_{X+x} +2 \int_0^{\infty} dx^d f_x f_{X-x}  \\
= \ & 2\int_{-\infty}^0 dx^d f_{-x} f_{X-x} + 2\int_0^{\infty} dx^d f_x f_{X-x}  \
= \ 
2\int_{-\infty}^{\infty} dx^d f_x f_{X-x} ,
\end{aligned}
\end{equation}
where we let $f_{(x_1, \dots, -x_i, \dots, x_d)} = f_{(x_1, \dots, x_i, \dots, x_d)}$ for $\forall i \in [1, d]$ (this also leads to $f_x = f_{-x}$).
In the first step we have used
\begin{equation}
\sum_{x_1 = 0}^\infty \dots \sum_{x_d = 0}^\infty f_{(x_1, \dots , x_d)} f_{X + (x_1, \dots , x_d)} = 
\sum_{x_1 = X_1}^\infty \dots \sum_{x_d = X_d}^\infty f_{(x_1, \dots , x_d)} f_{(x_1, \dots , x_d) - X}
\end{equation}

Finally, Eq.~\eqref{eq:MasterEq} becomes
\begin{equation}
\partial_t f_X = \theta \nabla^2 f_X - c f_X + \int_{-\infty}^{\infty} dx^d f_x f_{X-x}.
\end{equation}

Assuming the system already reaches steadiness, the left hand side vanishes as $\partial_t f_X = 0$. 
We first consider a simpler case of $\theta = 0$, i.e.,
\begin{equation}
0 = - c f_X + \int_{-\infty}^{\infty} dx^d f_x f_{X-x} .
\end{equation}
The ansatz for solving this equation is $f_x \sim 1/|x|^{d+1}$.

We turn back to consider the case of nonzero $\theta$.
It is clear to see  that the diffusion term $\theta \nabla^2 f_X \sim 1/|X|^{d+3}$ has the order much lower than the other terms $\sim 1/|X|^{d+1}$ in the differential equation.
Therefore, it is reasonable to ignore the diffusion term when investigating the asymptotic behavior of $f_x$ at large $x$.
Based on this, we argue that the solution of the nonlinear master equation in d+1 dimension has the asymptotic form $f_x \sim 1/|x|^{d+1}$ at large $x$.
Importantly, the solution also implies that the existence of a chaotic unitary background is not necessary for obtaining the steady state with special power-law correlations.

Then we discuss the entanglement structure of the steady state. From the nonlinear master equation we have obtain the distribution of the squared two-point correlation function, for mutual information between two points, it should give the same scaling, i.e. $I_{\rm{points}} \propto 1/r^{d+1}$. 
In particular, the $1/r^{d+1}$ correlation has a simple geometric interpretation in the quasiparticle picture.
This scaling form is  geometrically constrained by the “arc-length distribution” (see below) of the correlated points, 
which is the consequence of the dynamical balance of annihilation and creation of the quasiparticle pairs.
In quasiparticle picture, $I_{\rm{points}}$ corresponds to the probability that two given points with distance $r$ are connected. 
The integral over the surface area of these two points (a circular region with radius $r$) leads to $P_{\rm{arc}}(r) \propto 1/r^2$ that corresponds to the  probability that a given arc has length $l$. 
This is consistent with the quasiparticle picture, where the ``arc-length distribution'' $P_{\rm{arc}}(r)$ is argued to be scaled in $\propto 1/r^2$.
The consistency implies that the quasiparticle picture would work in any spatial dimension.
We now turn to derivation of the entanglement entropy. It can be calculated from the following integral in $d$ spatial dimension
\begin{equation}
S \sim \int_A dV_A \int_B dV_B P_{\rm{point}}(r_{A,B}) = \int_A dV_A \int_B dV_B \frac{1}{r_{A,B}^{d+1}} ,
\end{equation}
where $A$ is a disc/strip and $B$ is the complement of the system.
This integral leads to the asymptotic behavior $L_A^{d-1} \log L_A$ at the limit $L_A \to \infty$.

\end{widetext}

\bibliography{FreeFermion}

\end{document}